\documentclass[letter]{emulateapj}

\begin{document}

\title{XMM-NEWTON Observations of Ultraluminous X-Ray Sources in Nearby
Galaxies}

\author{Hua Feng and Philip Kaaret}

\affil{Department of Physics and Astronomy, The University of Iowa, Van Allen Hall
Iowa City, IA 52242; hua-feng@uiowa.edu}

\shortauthors{FENG \& KAARET}
\shorttitle{ULTRALUMINOUS X-RAY SOURCES WITH XMM-NEWTON}

\begin{abstract}

We examined X-ray spectral and timing properties of ultraluminous X-ray
sources (ULXs) in nearby galaxies in \textit{XMM-Newton} archival data.
There appear to be three distinct classes of spectra.  One class shows
emission from hot, diffuse plasma.  This thermal emission is similar to
that seen from recent supernovae; the temperatures are in the range
0.6--0.8~keV and the luminosities are the lowest in our sample, near
$10^{39}$~erg/s.  Three sources have spectra which are strongly curved
at high energies and have the highest temperatures in our sample,
1.0--1.4~keV.  These spectra are well fitted with a power-law plus
multicolor disk blackbody model with the power-law dominant at low
energies or a Comptonization model.  The remainder of the sources are
best fitted with a power-law plus multicolor disk blackbody model, as
is commonly used to describe the spectra of accreting black holes. 
These sources have the lowest thermal component temperatures,
0.1--0.4~keV, and extend to the highest luminosities, above
$10^{40}$~erg/s.  The temperature of the thermal component is in three
distinct ranges for the three source classes.  This diversity of
spectral shapes and the fact that the sources lie in three distinct
temperature ranges suggests that the ULXs are a diverse population. 
Two ULXs which show state transitions stay within a single class over
the course of the transition.  However, we cannot conclude with
certainty that the classes represent distinct types of objects rather
than spectral states of a single population of objects.  More
monitoring observations of ULXs with \textit{XMM-Newton} are required. 
We also searched for timing noise from the sources and report detection
of noise above the Poisson level from five sources.  In three of the
sources, the power density spectrum increases with decreasing frequency
as a power-law down to the lowest frequencies observed, below
$10^{-4}$~Hz.

\end{abstract}


\section{Introduction}

Non-nuclear ultraluminous X-ray sources (ULXs) in external galaxies
have been known since the \textit{Einstein} era \citep{fab89} with
apparent isotropic luminosities above the Eddington limit of a
10$M_\sun$ black hole ($L \approx 2 \times 10^{39}$ erg/s) \citep[for
reviews see][]{fab03,mil04b}.  From the surveys of nearby galaxies
undertaken with the \textit{ROSAT} HRI \citep{col02} and the
\textit{Chandra} ACIS \citep{swa04}, more than one hundred sources 
with luminosity greater than $10^{39}$ erg/s have been located in
nearby galaxies.

The nature of ULXs is still unclear.  The major dispute arises from our
observational ignorance of the angular distribution of the source
emission.  If X-rays are emitted isotropically, then the high
luminosity indicates that the ULXs might harbor intermediate mass black
holes (IMBHs) with the mass in the order of $100M_\sun$ or even higher
\citep{col99,mak00,pta99,kaa01}.  If the emission is beamed, then the
ULXs could be stellar mass black holes \citep{kin01,kor02}.

Evidence that the luminosities are truly high for some ULXs has been
found from their associations with optical nebula which show
indications of X-ray ionization \citep{pak03}.  The photoionizing flux
required to power the nebula surrounding one ULX requires a luminosity
greater than $4 \times 10^{39}$ erg/s \citep{kaa04}.  Optical nebulae
associated with some ULXs also show shock emission indicative of
unusually powerful explosions \citep{pak03}.

X-ray spectral analysis has revealed thermal components in the spectra
of several ULXs with temperatures of 0.1-0.3~keV
\citep{col99,kaa03,mil03}, which are lower than that in stellar mass
black holes in high states, suggesting accretion onto a black hole with
higher mass \citep{mak00}.   Also, a few ULXs exhibit breaks in their
timing power spectra at relatively low frequencies which have been
interpreted as evidence for high black hole masses
\citep{cro04,sor04}.  However, such interpretations are ambiguous due
to our lack of understanding of timing power spectra and the fact that
some stellar-mass black holes occasionally exhibit breaks at similar
frequencies \citep{mcc04}.

\textit{XMM} is more powerful for spectral and timing analysis than
previous X-ray telescopes owing to its large effective area.  The goal
of this paper is to use \textit{XMM} archival data to study the
spectral and timing behaviors of ULXs in nearby galaxies.  In
\S~\ref{sec:data} we describe the details of data reduction.  We
present the spectral analysis and details for some selected ULXs in
\S~\ref{sec:spec} and timing analysis and results in
\S~\ref{sec:timing}.  Finally, we interpret our results in
\S~\ref{sec:discussion}.

\section{Data reduction}\label{sec:data}

We derive our source sample from previously reported ULX detections. 
Specifically, we used the \textit{Chandra} catalog of \citet{swa04},
the \textit{ROSAT} catalog of \citet{col02}, and the catalog of
\citet{liu05}.  We included only sources located in nearby galaxies
with distances less than 11~Mpc. We selected sources with enough counts
in \textit{XMM} exposures for good spectral and timing analysis.  We
eliminated sources crowded near other point sources that can not be
resolved by \textit{XMM}, e.g.\ the ULXs in M82.  Starting from these
three catalogs and applying the criteria above, which are solely
intended to enable us to perform reliable spectral and timing analysis,
we produced a final sample of 28 sources, see Table \ref{tab:data}. 
Hereafter we refer to sources by their number in Table \ref{tab:data}
since not every ULXs has a common name.  Our sample is not meant to
constitute an unbiased survey of nearby galaxies for ULXs. Rather, it
is intended to provide a sample with which we can examine the spectral
and timing properties of sources previously reported as ULXs in the
literature.

All data were processed using \textsl{SAS} 6.1.0 with the most recent
calibration files as of March 2005.  We extracted single and double
events (PATTERN $\leq$ 4) for pn data with the addition of triple
events (PATTERN $\leq$ 12) for MOS data.  For spectral analysis, we
required FLAG $=$ 0 and screened for good time intervals without
background flares.  For timing analysis, we required FLAG equal to
\#XMMEA\_EM for MOS and \#XMMEA\_EP for pn, and selected a continuous
good time interval without background flare contamination or data
gaps.  We found that gaps in the \textit{XMM-Newton} data stream often
cause spurious timing signals in the frequency region $10^{-4} \sim
10^{-2}$ Hz.  Therefore it is better to avoid gaps than attempt to fill
them, e.g.\ with an average count rate.  

Source counts are generally extracted from a 32\arcsec-radius  circle,
following the recommendation in the data analysis threads on the XMM
SAS web page at
\url{http://xmm.vilspa.esa.es/sas/documentation/threads/}.  For  some
very bright isolated sources, we used a 40\arcsec-radius circle (this
results in an increase in luminosity by a few percents but no
difference in spectral parameters within errors).  For sources in
crowded regions or with a strong diffuse background, we used a source
region with a 16\arcsec-radius circle.  To attain the best statistics,
we combine the data from the pn and the two MOS detectors unless some
CCD detectors are turned off, the source rides on the gap between two
CCD chips in one of the detectors, or the source is located at the edge
of a CCD chip.  The energy spectra are extracted from individual MOS
and pn detectors and then grouped for fitting.  The light curves are
obtained from the merged events files.

\section{Spectral analysis and results}\label{sec:spec}

The energy spectra are fitted with \textsl{XSPEC} 11.3.1.  First, we
fit all 28 sources with a single power-law spectrum and, separately,
with a single multicolor disk (MCD) \citep[][diskbb in
\textsl{XSPEC}]{mit84,mak86} spectrum.  Both spectral models, and all
others considered here, also include interstellar absorption.  The
absorption column density is treated as a free parameter with a lower
bound set to the column density along the line of sight within the
Milky Way \citep{dic90} for all models in this paper.  Then we tried
adding a MCD component to each power-law model.  We accept the MCD
component if its addition improves the fit by at least 4$\sigma$
significance as calculated using the F-test for an additional
component.  The spectral parameters from all the fits, the unabsorbed
0.3--10 keV luminosities calculated assuming isotropic emission, and
$\chi^2$/dof for each fit are listed in Table~\ref{tab:spec}.  We
include only those fits with $\chi^2$/dof $\le 1.4$.  We find 13
sources with a significant MCD component in addition to the the
power-law component.  In half of these sources, the power-law is
dominant at high energies and the MCD disk temperature is in the range
0.1-0.4~keV.

Source 1 is better fitted with a single MCD model than with a single
power-law model.  The F-test shows that the MCD model is preferred at 
a significant level of 0.05.  For sources 2, 11, 13--16 and 19--21, the
F-test shows no preference for a single MCD or a single power-law.  For
sources 9, 10, 18, 24, and 27, the $\chi^2$/dof for the single MCD
model is greater than 1.4, therefore that model is rejected, and no
significant improvement was obtained with the MCD plus powerlaw model. 
For these sources, only the powerlaw fit is reported.

\begin{figure}[b]
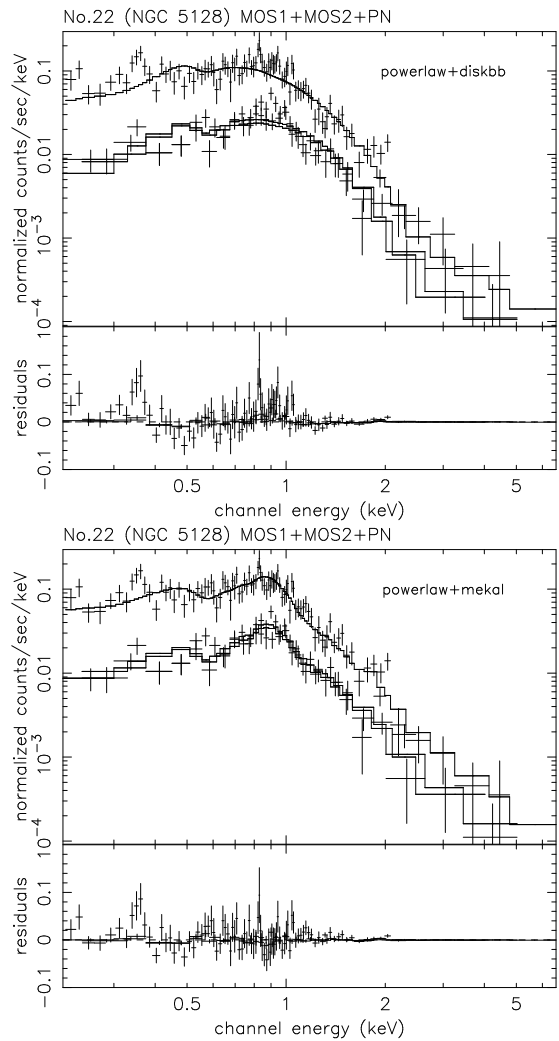

\centering
\includegraphics[width=2.7in,angle=-90]{f1a.eps}\\
\includegraphics[width=2.7in,angle=-90]{f1b.eps}
\caption{Energy spectra of source 22 fitted with an MCD model plus
power-law model (top) or with a mekal plus power-law model (bottom).
The mekal model is a significantly better to fit the spectrum because
it adequately models the bump around 0.5--1.0 keV.
\label{fig:mekal}}
\end{figure}

\begin{figure*}[t]
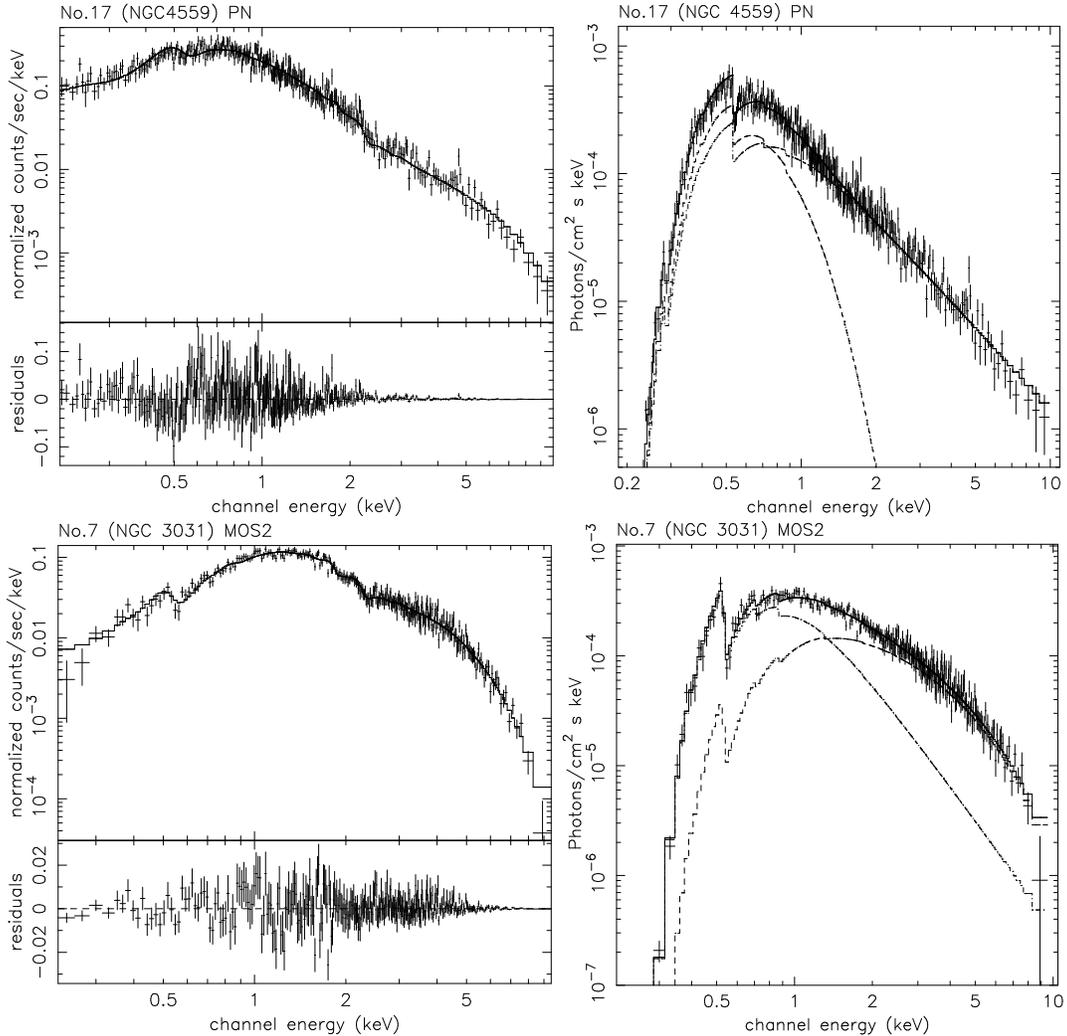

\centering
\includegraphics[width=2.7in,angle=-90]{f2a.eps}
\includegraphics[width=2.7in,angle=-90]{f2b.eps}\\
\includegraphics[width=2.7in,angle=-90]{f2c.eps}
\includegraphics[width=2.7in,angle=-90]{f2d.eps}\\
\caption{Energy spectra of sources 17 and 7.  The spectrum of source 17
(top left) consists of a 0.17 keV MCD and a power-law (top right). The
spectrum of source 7 is curved at high energy (bottom left) which
could be fitted either by a combination of a 1.51 keV MCD and a
power-law (bottom right) or by a single Comptonization model (not
shown).}
\label{fig:spec}
\end{figure*}

The spectra of sources 12, 22 and 26 show significant residuals when
fitted with a MCD plus power-law model and have $\chi^2/{\rm dof} >
1.7$.  Therefore, we searched for another spectral model which would
give a better fit.  All three spectra show a bump around 0.5--1.0 keV,
which is a region where hot diffuse gas produced significant line
emission.  We found that the sum of a hot diffuse gas model 
\citep[][mekal in \textsl{XSPEC}]{mew95} plus a power-law produced a
much better fit than the MCD plus power-law model for all three
sources.  A comparison of the two models for source 22 is shown in
Fig.~\ref{fig:mekal}.

For sources 3, 6, and 7, the MCD component is dominant at high
energies.  We fitted the spectra for these three sources with a single
MCD model and with a single Comptonization model \citep[][compst in
\textsl{XSPEC}]{sun80}.  The single MCD model fits are not poor, with
$\chi^2_{\nu} = 1.2 \sim 1.3$, but are rejected compared to the MCD
plus power-law model or the Comptonization model.  The Comptonization
model fits, with large optical depth $\tau\approx20$, are as good or
better than the power-law plus MCD model fits.  These three sources
shown significant curvature in the high energy parts of their spectra,
which is quite distinct from the power-law followed by most of the
other sources at high energies. Fig.~\ref{fig:spec} shows the spectrum
of source 17, which is best fitted with an MCD plus power-law with the
power-law dominant at high energies and which is typical of the sources
with low disk temperatures, $kT = 0.1-0.4$~keV.  In contrast, the
spectrum of source 7 is the strongly curved at high energies, as shown
fitted with an MCD plus power-law model in which the MCD component is
dominant at high energies or a Comptonization model.  The spectra of
sources 3 and 6 have the same shape as source 7.

\begin{figure}
\centering
\includegraphics[width=2.7in,angle=0]{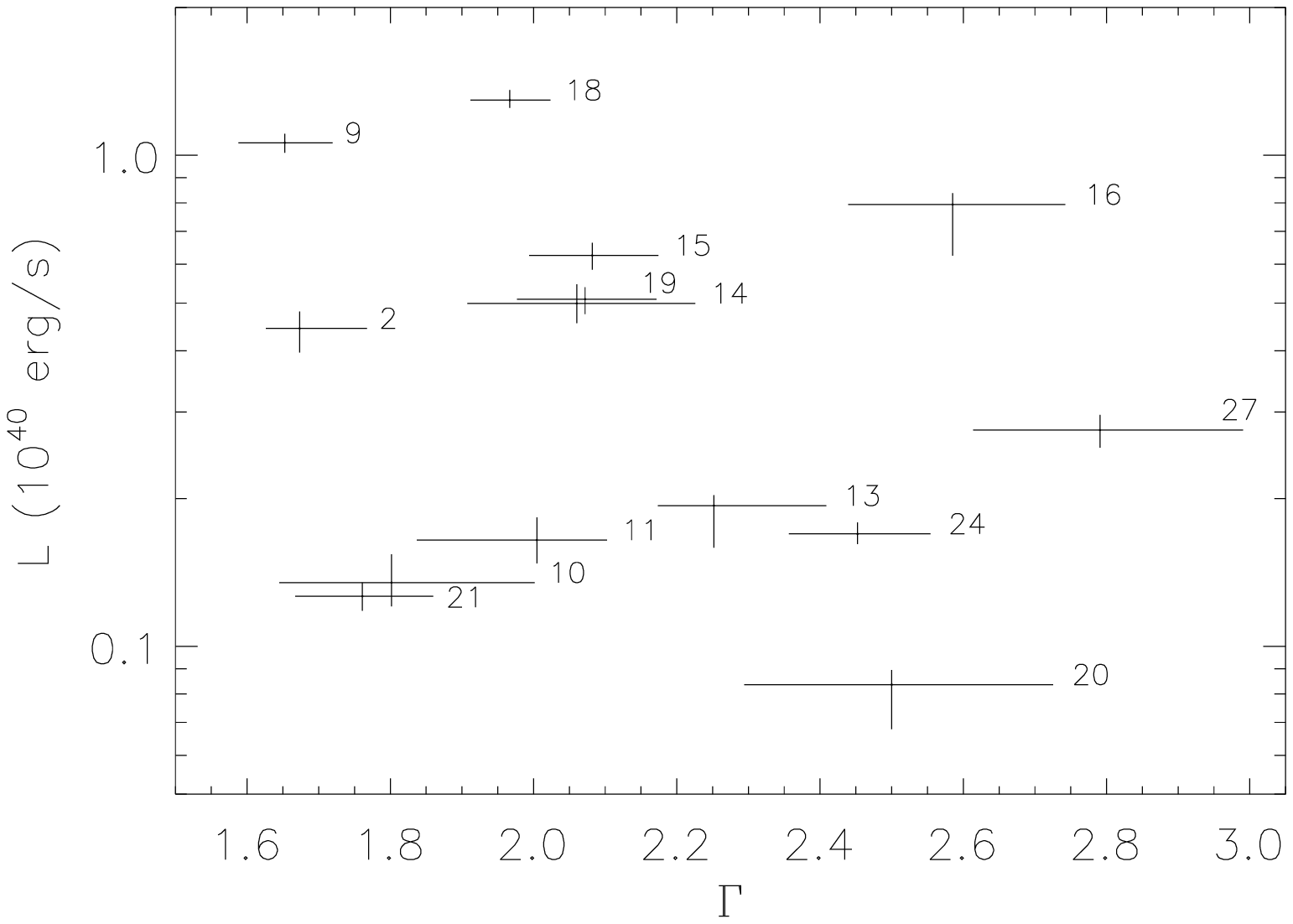}\\
\includegraphics[width=2.7in,angle=0]{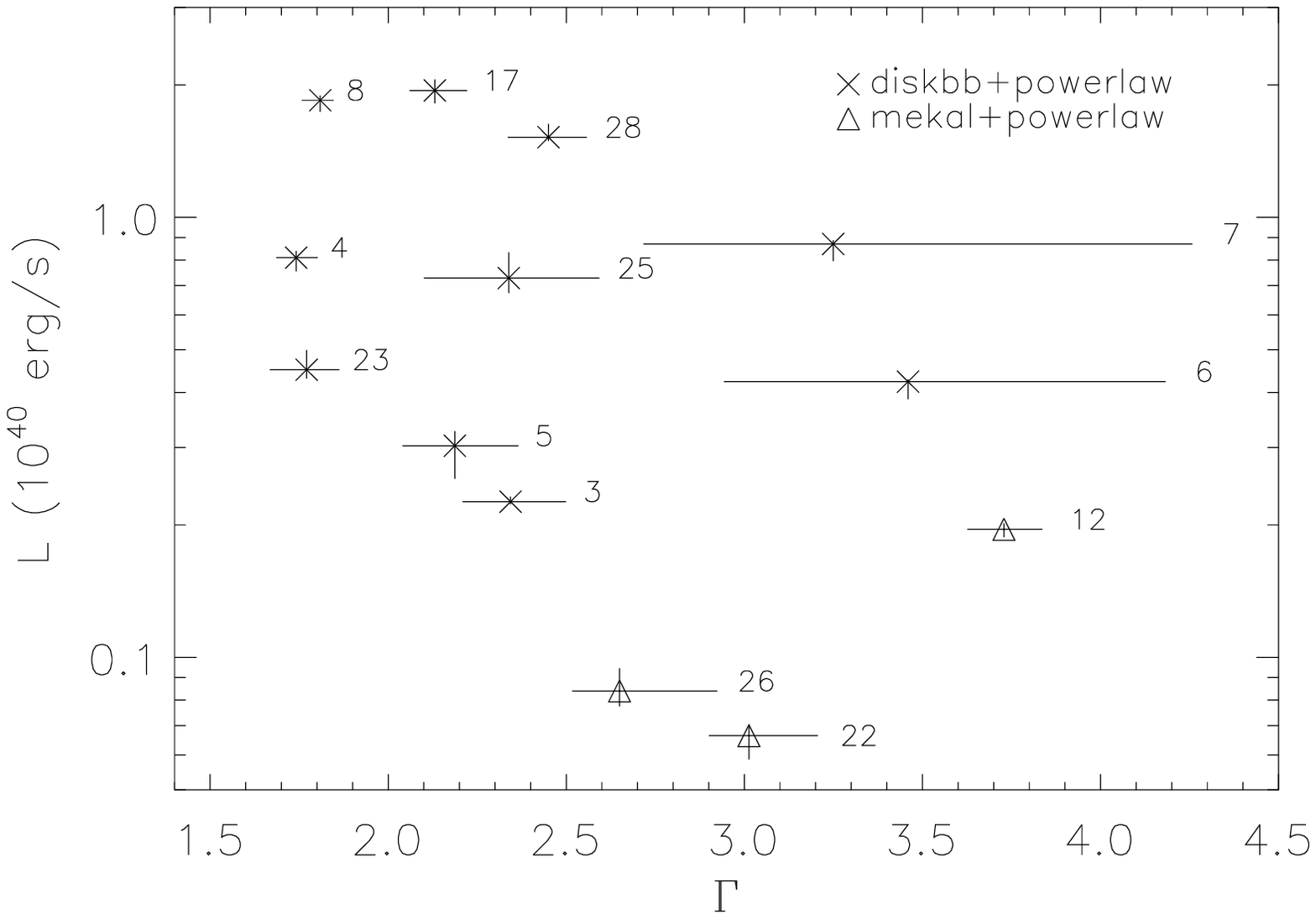}
\caption{Luminosity vs. photon index of ULXs. Top: ULXs with a single
power-law spectrum; Bottom: ULXs with a prominent MCD component plus the
power-law spectrum. The label indicates the source index number
in Table \ref{tab:data}.
\label{fig:lg}}
\end{figure}

\begin{figure}
\centering
\includegraphics[width=2.7in,angle=0]{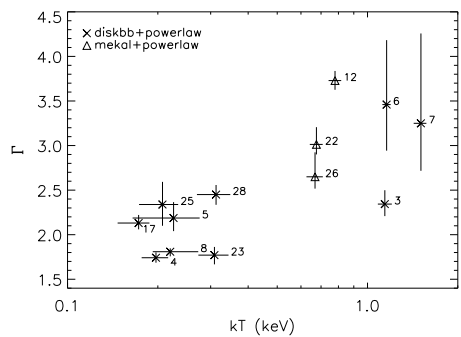}
\caption{Photon index vs. temperature of ULXs with a prominent thermal
component plus the power-law spectrum. The label indicates the source index number
in Table \ref{tab:data}.
\label{fig:gt}}
\end{figure}

\begin{figure}
\centering
\includegraphics[width=2.7in,angle=0]{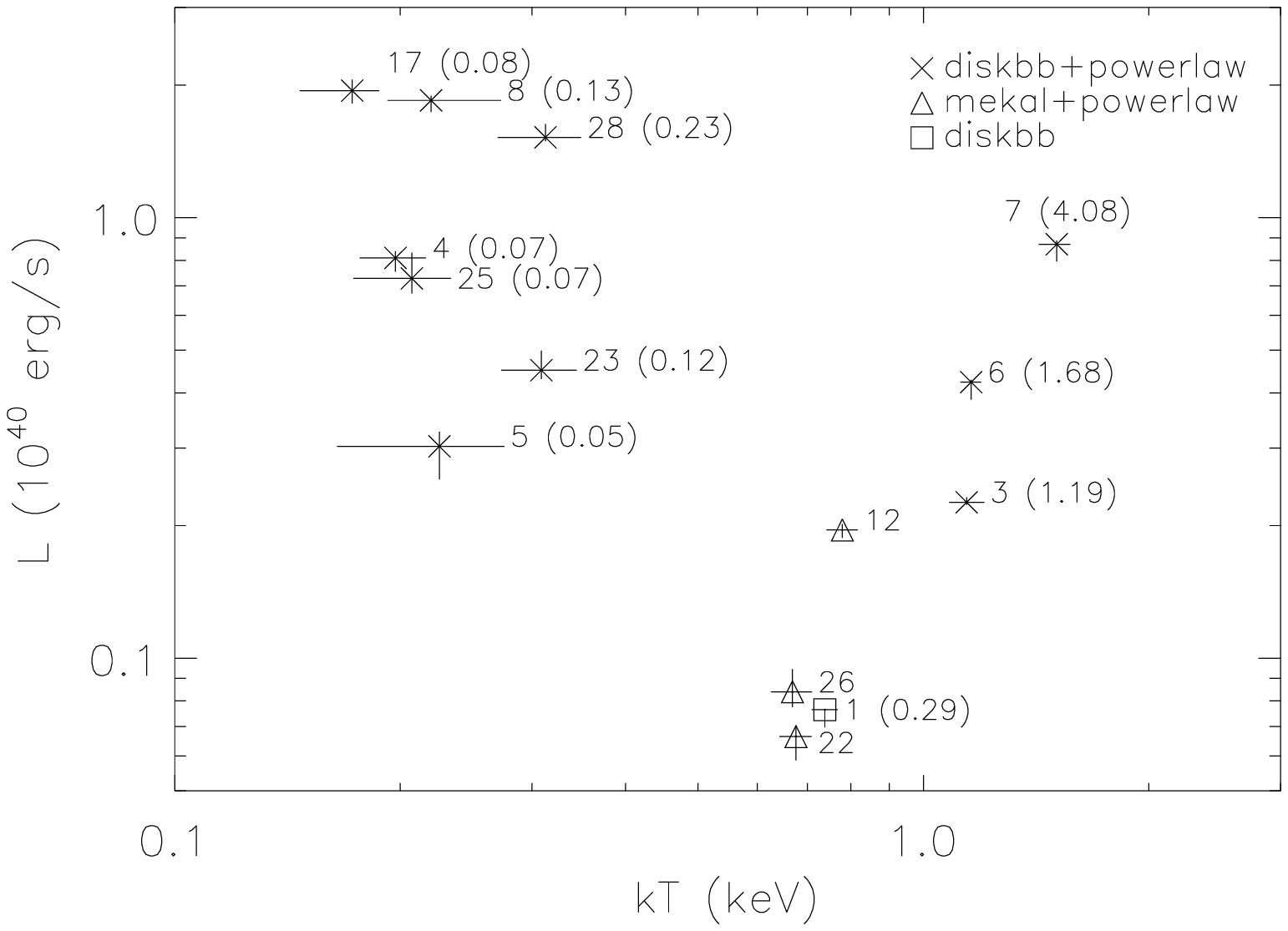}\\
\includegraphics[width=2.7in,angle=0]{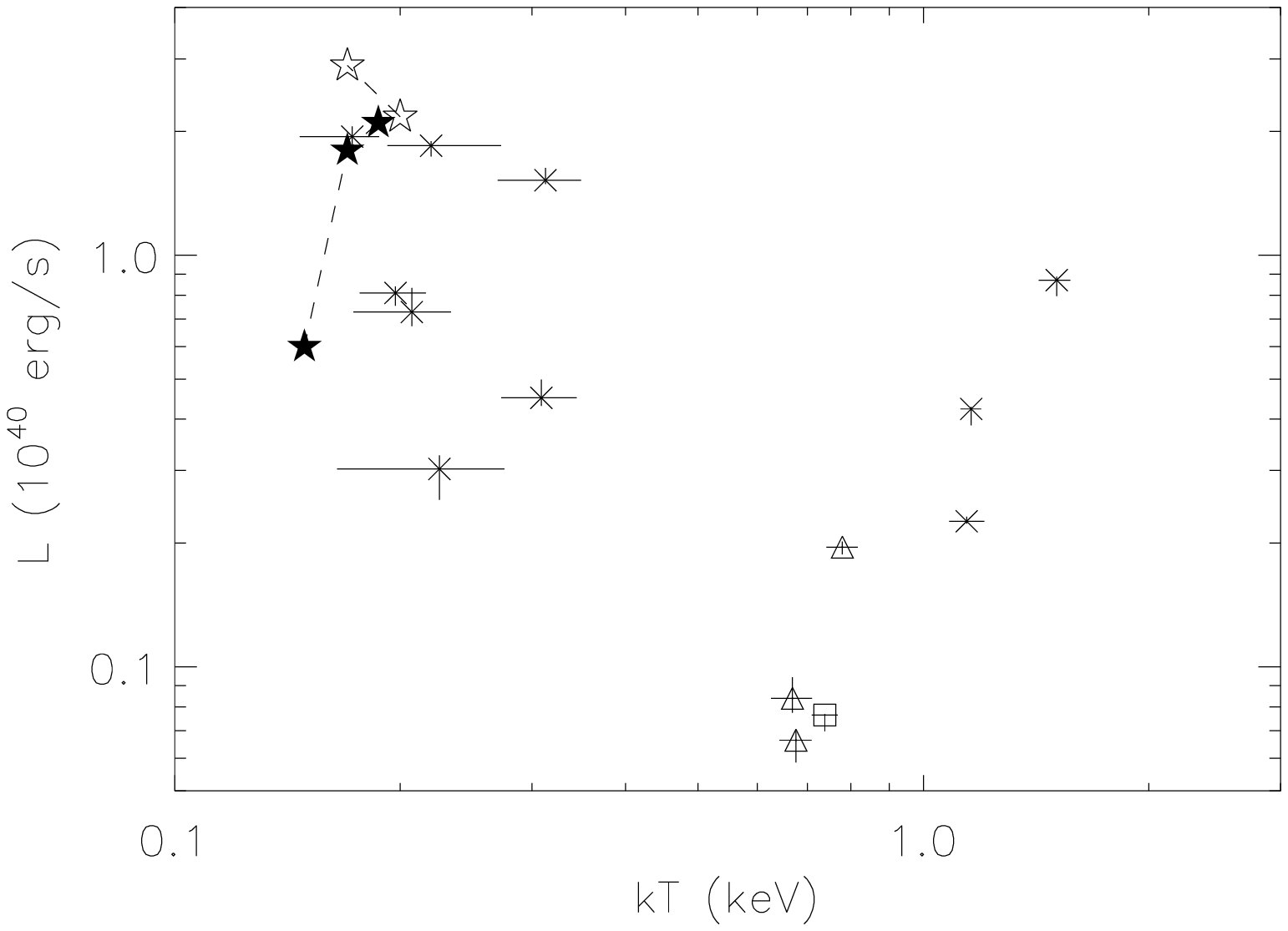}
\caption{Luminosity vs. temperature of ULXs with a significant thermal
component in their spectra.  The label indicates the source index
number in Table \ref{tab:data} and the fractional Eddington luminosity
in the brackets which is derived from the MCD model (for details, see
the text).  In the bottom panel, two ULXs which have been observed 
with \textit{XMM-Newton} to undergo state transitions are added.  The
filled stars represent Holmberg II X-1 \citep{dew04} and the white
stars represent Holmberg IX X-1 \citep{wan04}. The dashed lines connect
individual sources observed in different states.
\label{fig:lt}}
\end{figure}

In Fig.~\ref{fig:lg}, we plot source luminosity $L$ versus photon index
$\Gamma$ for sources with and without a significant thermal component.
For sources without a thermal component there is no  apparent
correlation between luminosity and photon index.  For sources with a
thermal component, there appears to be an absence of sources with hard
spectra at low luminosities, i.e. in the bottom left corner of the
bottom panel of Fig.~\ref{fig:lg}.

The relation between the photon index $\Gamma$ of the power-law
component and the temperature $kT$ of the thermal component is shown in
Fig. \ref{fig:gt} for all sources where addition of the thermal
component significantly improves the fit.  Considering all sources
together, high temperatures appear to be correlated with soft photon
indices.   However, this correlation appears to be a by-product of the
juxtaposition of different spectral shapes.  In particular, sources 3,
6 and 7 which have the highest temperatures are well fitted by a single
Comptonization model or a  power-law plus MCD model with the power-law
is dominant at low energies and the thermal component at high energies,
in contrast to the other sources.  The intermediate temperature
sources, 12, 22, and 26, are exactly those better fitted by a mekal
model than a MCD model. Considering only the sources for which the
power-law plus MCD is the best fit and the power-law dominates at high
energies, there is no correlation apparent between $\Gamma$ and $kT$.

Fig.~\ref{fig:lt} shows the relationship between the temperature $kT$
of the thermal component and the X-ray luminosity $L$, for all sources
where addition of the thermal component significantly improves the fit.
Again, the sources segregate according to the best fitting spectral
model.  Sources 3, 6 and 7 have the highest temperatures and cover a
wide luminosity range extending up to $10^{40}$~erg/s.  The sources
with mekal components or a single MCD are at intermediate temperatures
and relatively low luminosities. The sources for which the power-law
plus MCD is the best fit tend to have low temperatures and relatively
high luminosities.  There is no correlation apparent between $L$ and
$kT$ if we consider only the latter sources.

We can, crudely, estimate the ratio of the observed luminosity to the
Eddington luminosity using the fitted temperature and X-ray luminosity.
By adopting the canonical values of the color correction and
temperature profile correction in \citet{mak00}, the disk blackbody
temperature scales with black hole mass as

\begin{equation} kT=1.2 \beta^{-\frac{1}{2}} \eta^\frac{1}{4}
\left(\frac{M}{10M_\sun}\right)^{-\frac{1}{4}} \quad \mathrm{keV},
\end{equation}

\noindent where $\beta$ is determined by the black hole spin, $\beta =
1$ for a Schwarzschild black hole and $\beta=1/6$ for a maximally
rotating Kerr black hole, and $\eta$ is the bolometric luminosity
normalized to the Eddington luminosity as $\eta =
L_\mathrm{bol}/L_\mathrm{Edd}$.  We take $L_\mathrm{Edd} = (1.3 \times
10^{38} \mathrm{erg/s}) ~ (M/M_\sun)$ and approximate $L_\mathrm{bol} =
L_\mathrm{X}$.   Solving for $\eta$, we find the fractional Eddington
luminosity

\begin{equation} \eta = \beta
\left(\frac{kT}{1.2\;\mathrm{keV}}\right)^2
\left(\frac{L_\mathrm{X}}{1.3\times10^{39}~\mathrm{erg/s}} \right)^{1/2}.
\label{eq_fLedd} \end{equation} 

We derive a fractional Eddington luminosity (shown in Fig.~\ref{fig:lt}
in the brackets) from Eq.~\ref{eq_fLedd} with $\beta=1$.  There are
three sources which appear to be above the Eddington limit, with
$L/L_\mathrm{Edd}$ of 1.2, 1.7 and 4.1 respectively for sources 3, 6,
and 7.  If the black holes are rapidly rotating, as suggested by
\citet{mak00}, then the sources fall below Eddington luminosity with
source 7 still near the Eddington luminosity.

\section{Timing analysis and results}\label{sec:timing}

\begin{figure}
\centering
\includegraphics[width=2.7in]{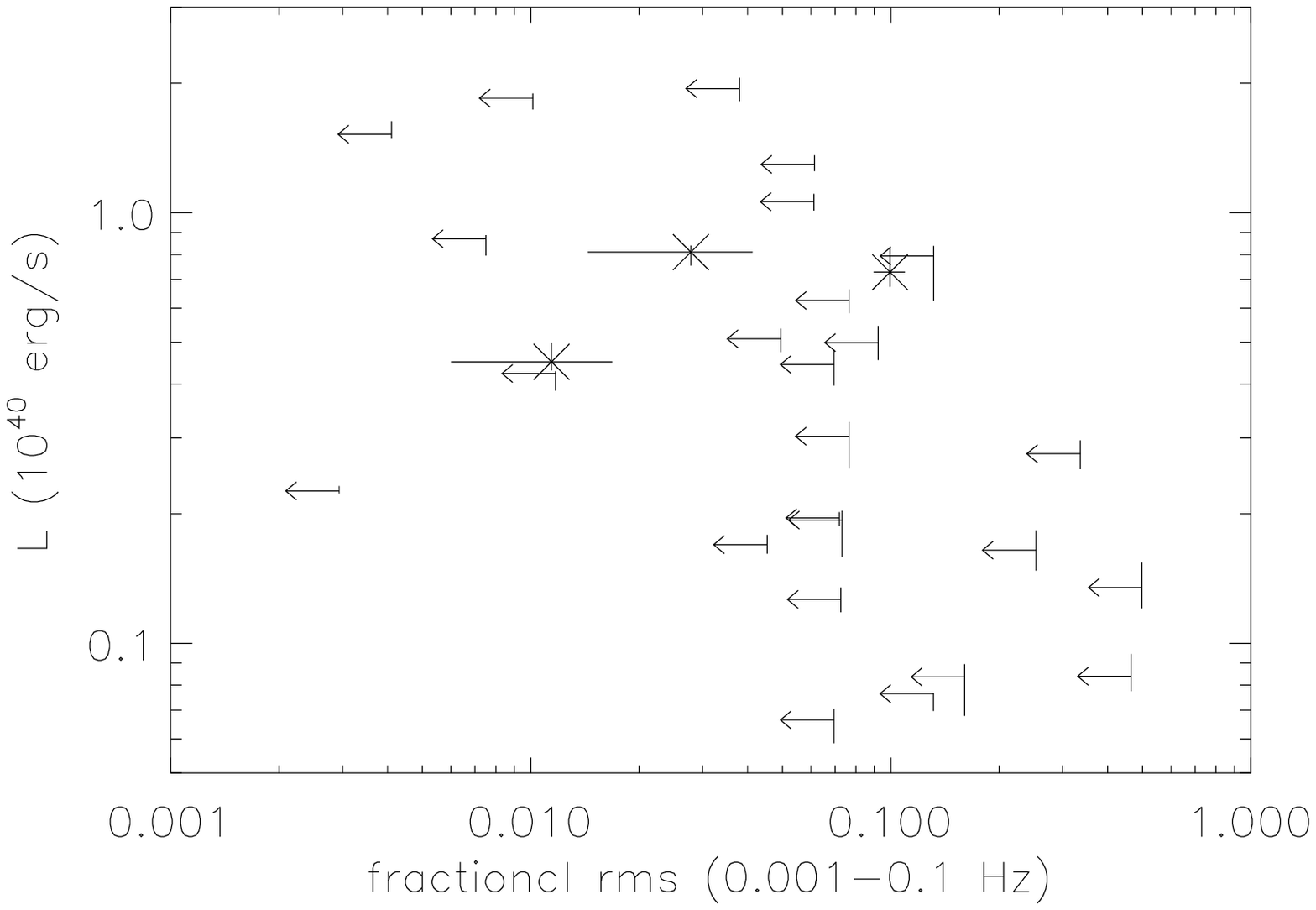}\\
\includegraphics[width=2.7in]{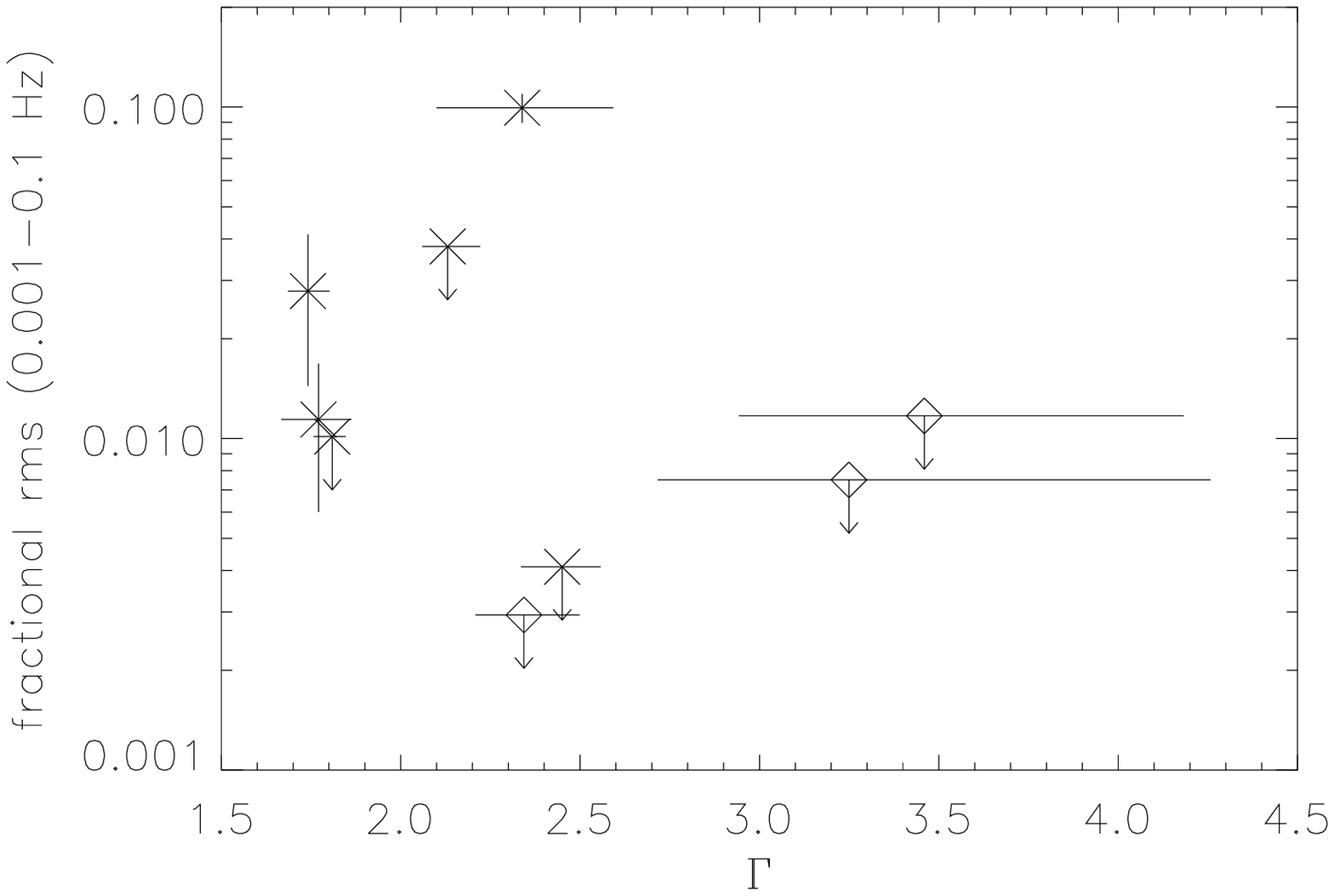}\\
\includegraphics[width=2.7in]{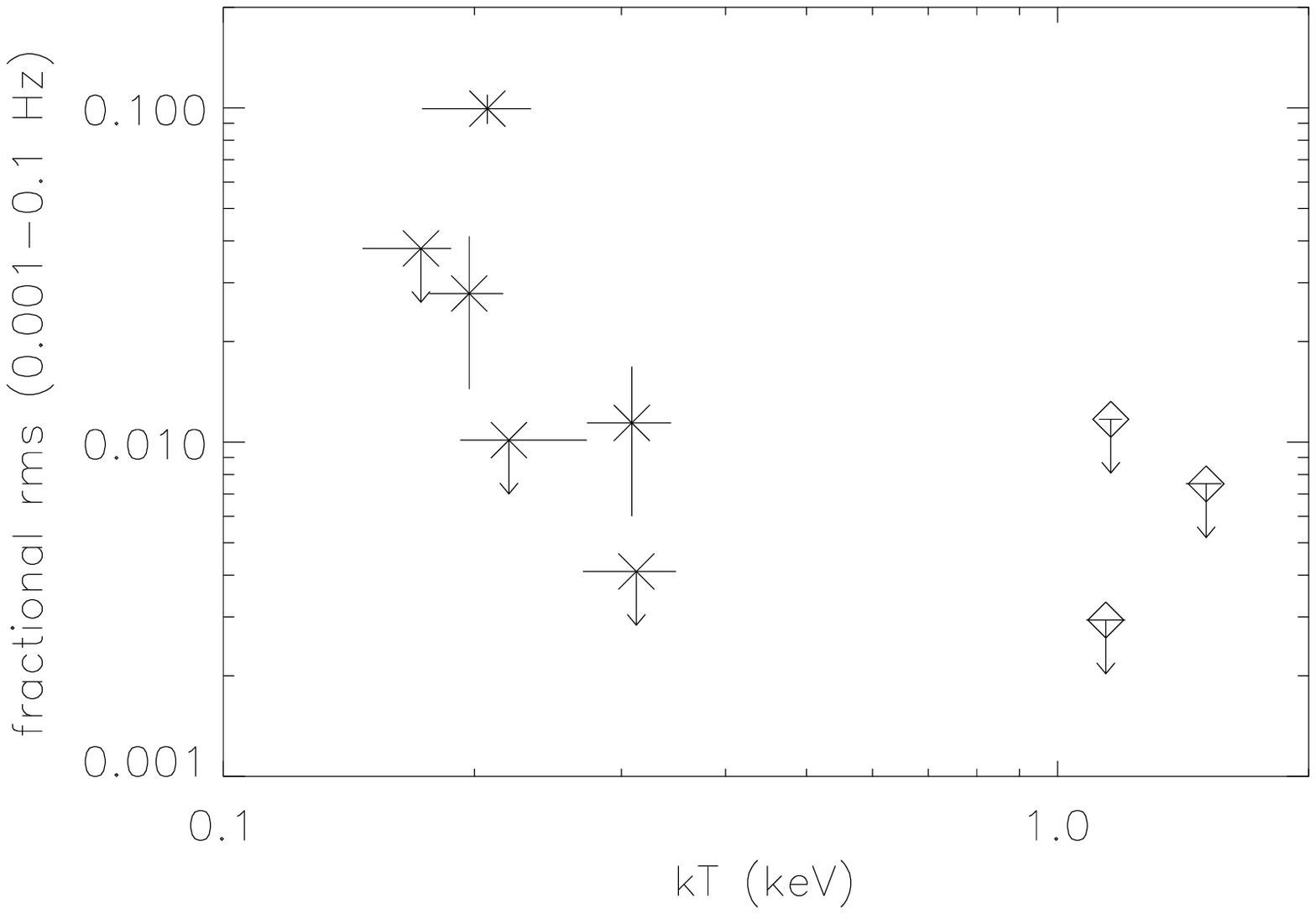}
\caption{The relationship between the fractional rms integrated in the
frequency region 0.001--0.1 Hz and the X-ray luminosity (top), power-law
photon index (middle) and the thermal temperature (bottom). The
luminosity is taken from the best-fitted spectrum. Three sources have
detectable timing noise; these sources are best fitted with a MCD plus 
power-law spectrum.  The sources for which
timing noise above the Poisson level is not detected are shown as
2-$\sigma$ upper limits.  In the middle and bottom panels, only sources
which are well fitted with a MCD plus power-law model are plotted. Sources
without detected noise are plotted only if the upper limit on the rms
is less than 0.05.  The diamonds indicate sources which were well
fitted with a Comptonization model or a high temperature MCD plus
power-law model.
\label{fig:rms}}
\end{figure}

After examining the power spectral density (PSD) of every source in our
sample, we find three sources (No. 4, 23, \& 25) with detectable
noise power in the frequency range 0.001--0.1~Hz.  The lower frequency
bound is chosen based on the shortest good time interval of any
observation used in this analysis in order to allow a uniform frequency
range for all sources.  The PSD in the other sources is consistent with
Poisson fluctuations, and we can only place an upper limit on the noise
power (see Fig. \ref{fig:rms}).  There is no obvious correlation
between the rms variability and the luminosity among those three sources
with detectable noise power.  The rms upper limits appear correlated
with source luminosity because the sensitivity of the upper limit
depends on the total number of counts recorded.

\begin{figure*}
\centering
\includegraphics[width=2.7in]{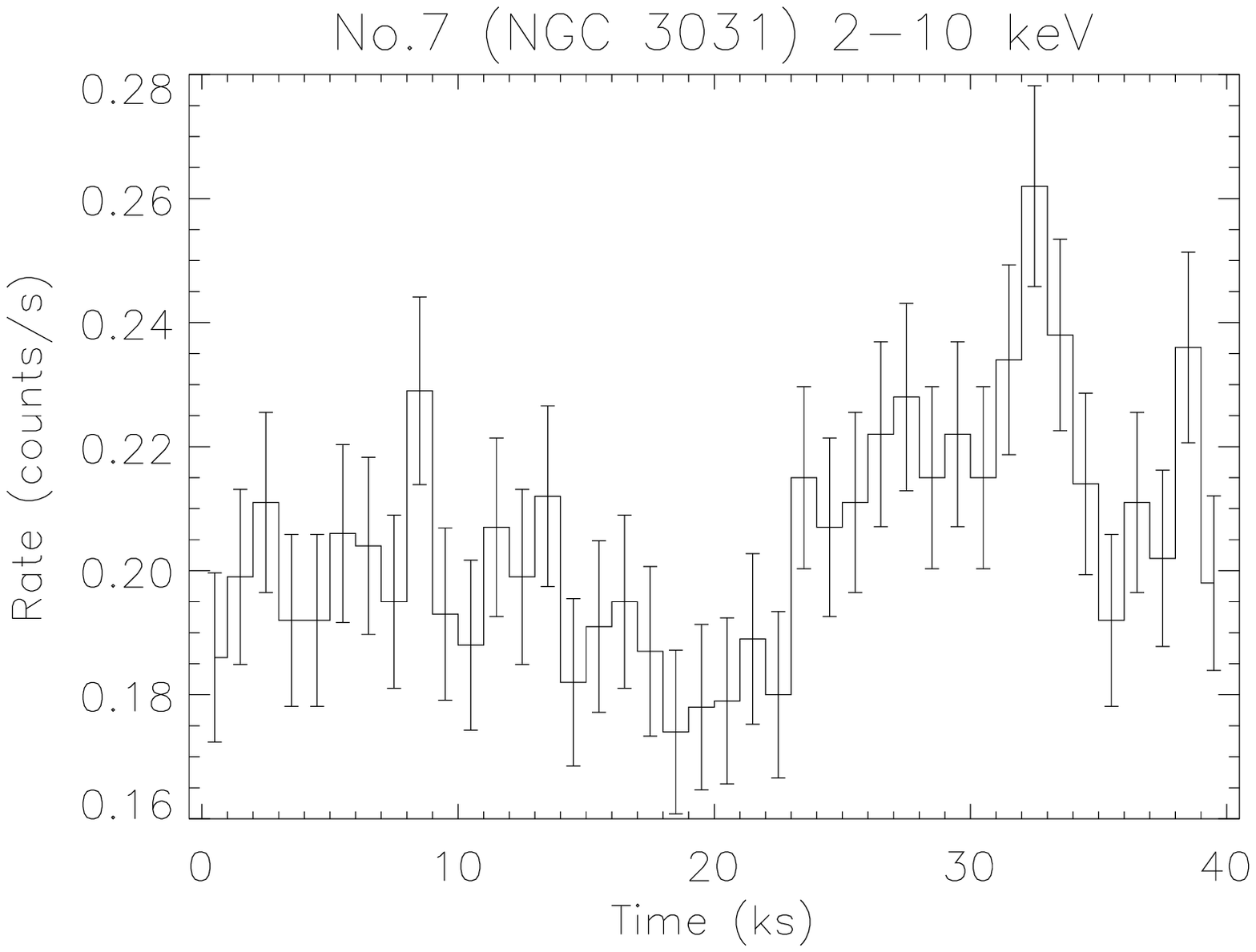}
\includegraphics[width=2.7in]{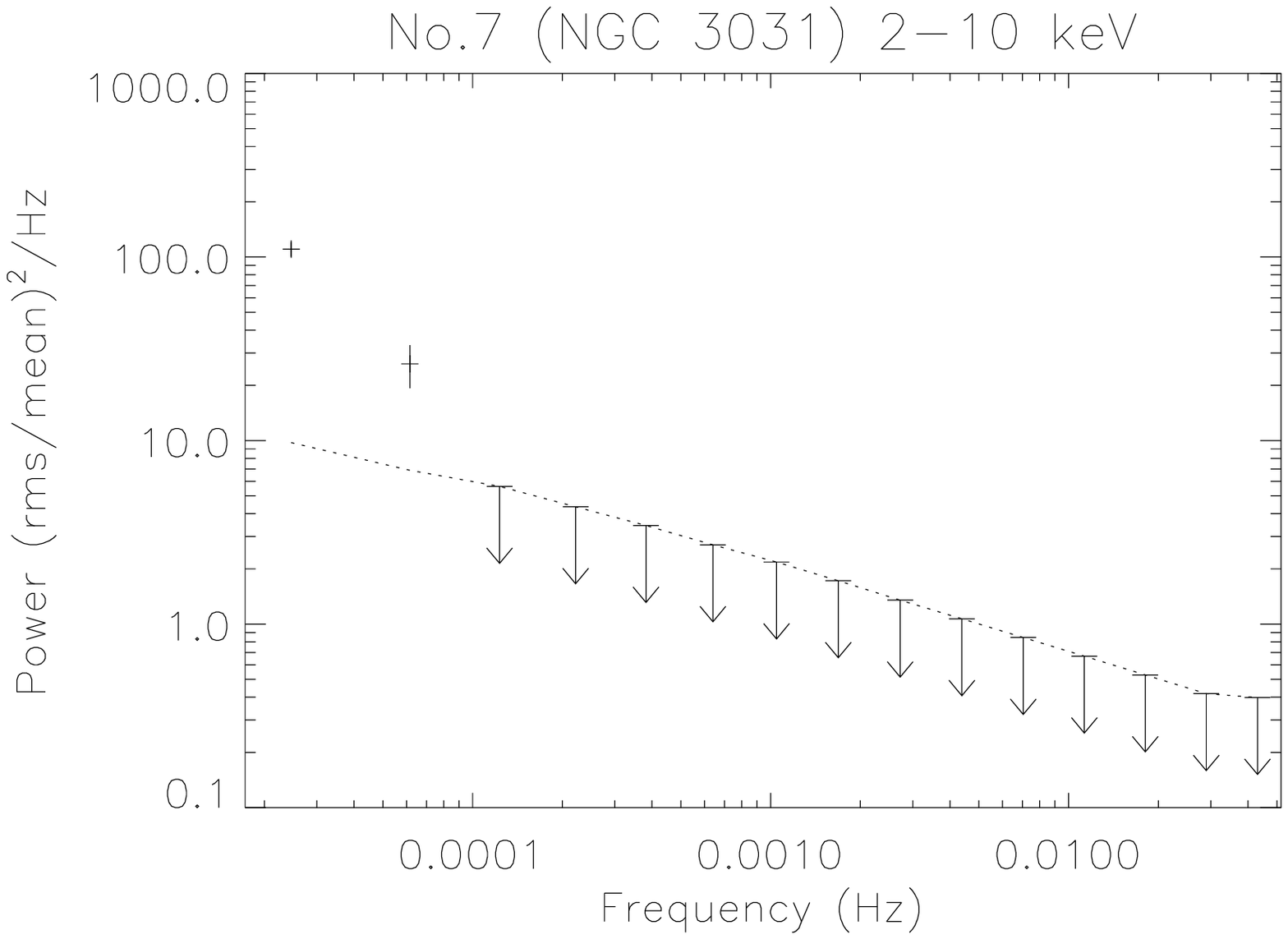}\\
\includegraphics[width=2.7in]{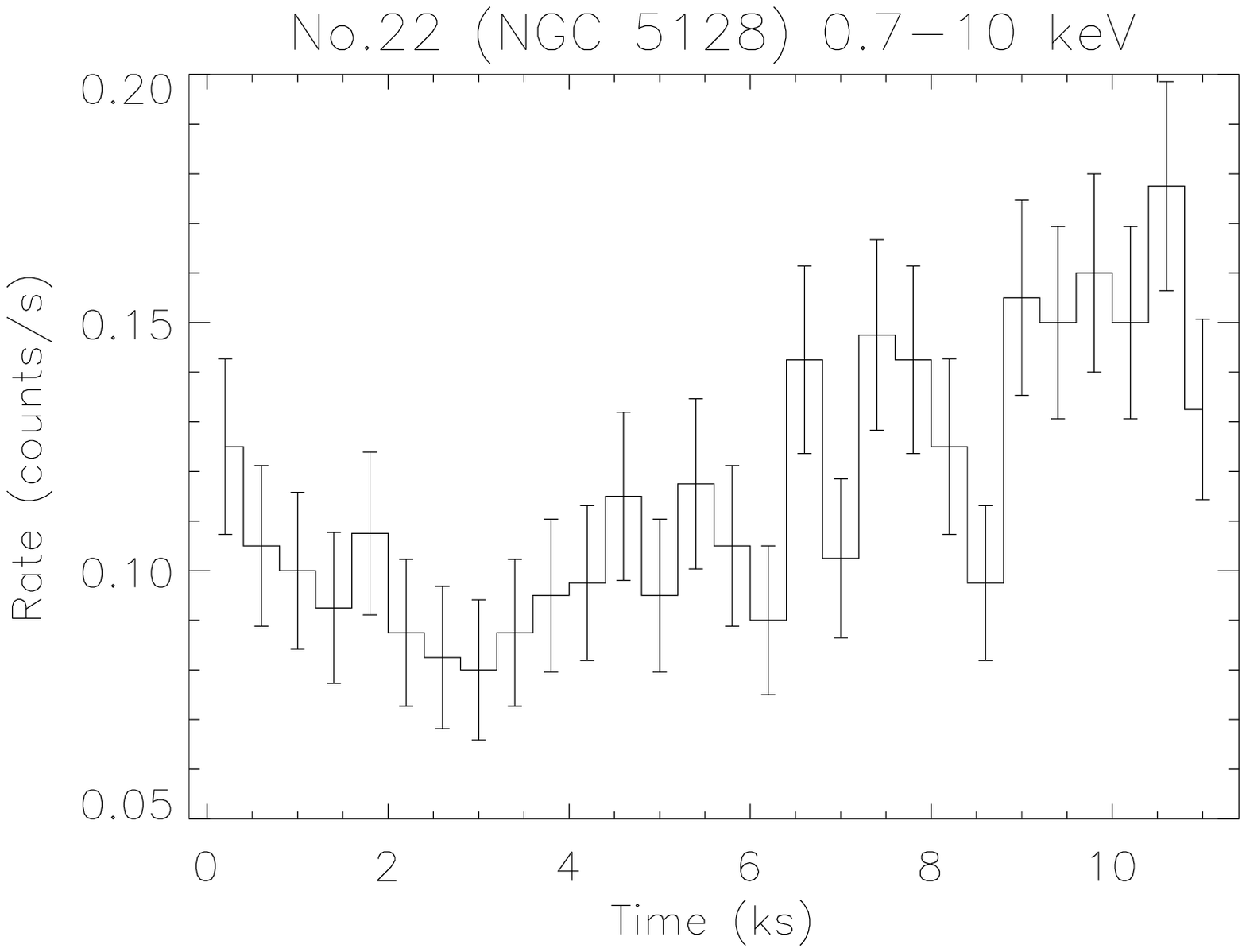}
\includegraphics[width=2.7in]{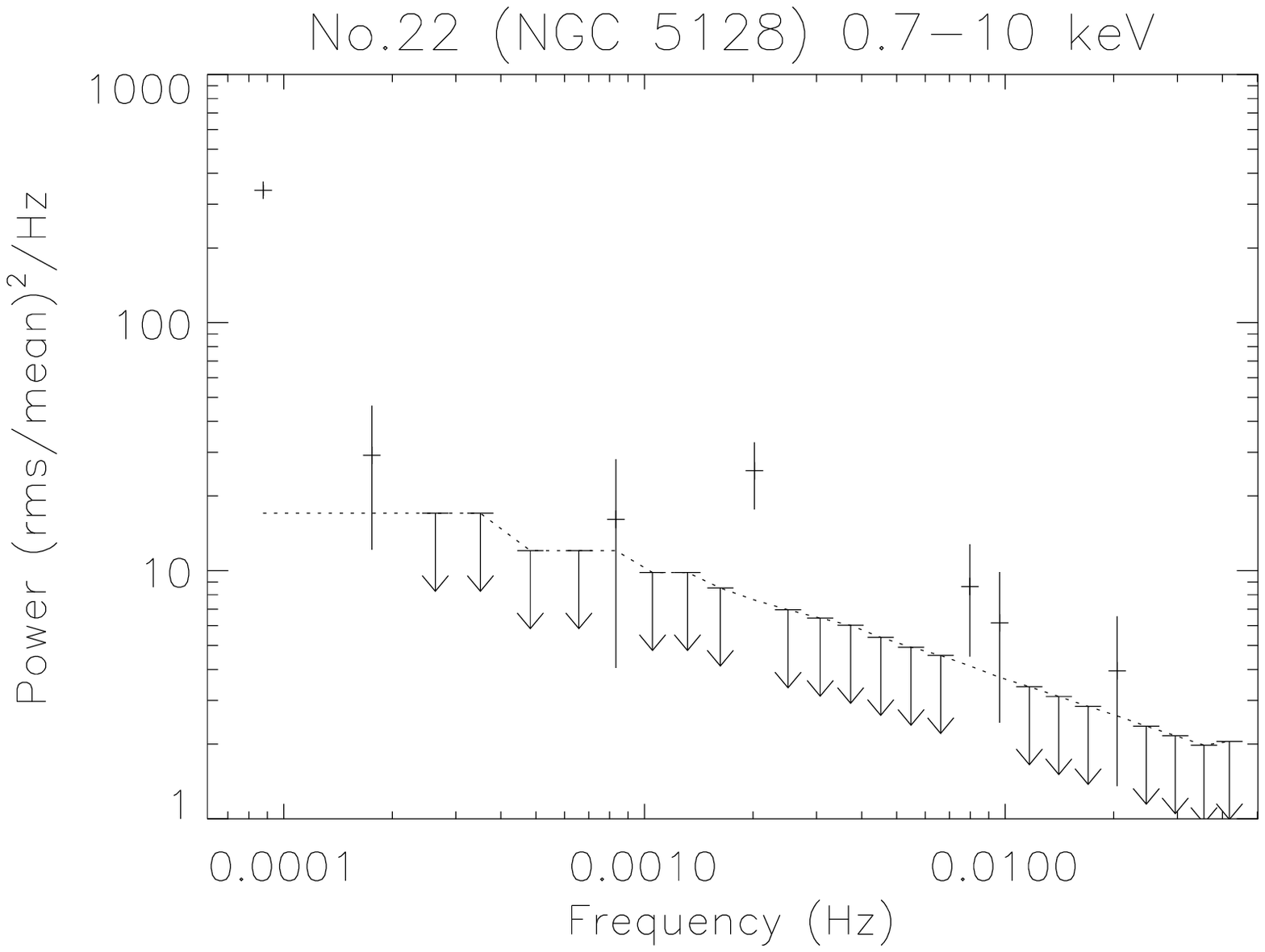}\\
\includegraphics[width=2.7in]{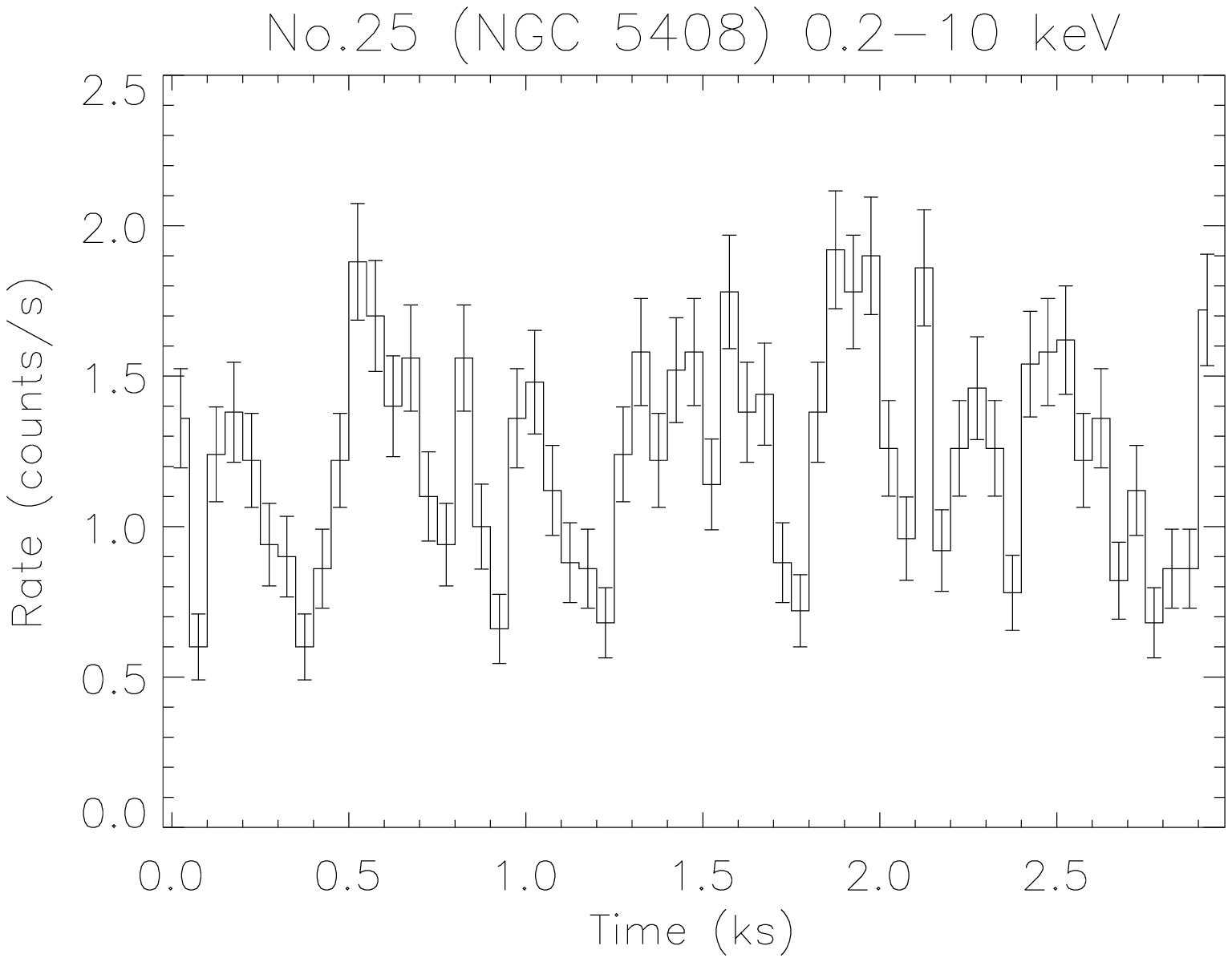}
\includegraphics[width=2.7in]{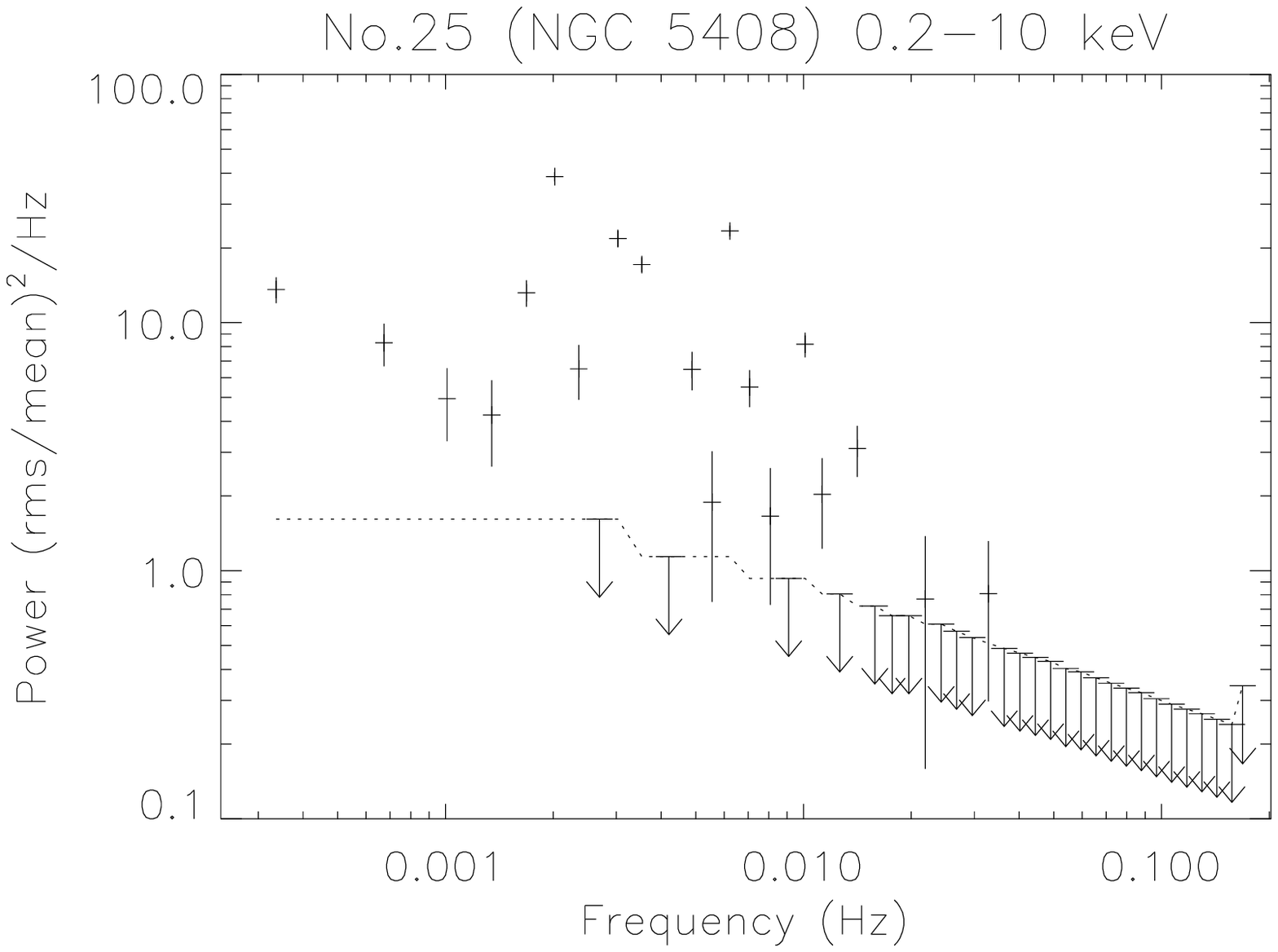}\\
\caption{Light curves and power spectra of three ULXs.
\label{fig:psd}}
\end{figure*}

We are able to detect timing noise in several frequency bands for three
sources, No.~7, 22, \& 25.  Light curves and PSDs for these three
sources are presented in Fig.~\ref{fig:psd}.  All PSDs are normalized
to rms \citep{van88} and binned in logarithmic scale with points above
the Poisson level plotted with error bars and points below the Poisson
level plotted as an upper limit.  We have checked the PSD from
background regions for these observations to ensure that the timing
noise is not produced by variations in the background.  The noise power
for sources 7 and 22 lies below 0.001 Hz.  They do not have detectable
noise power in 0.001--0.1~Hz band used above.

We note that the timing features reported by \citet{cro04}, including
PSDs with a break power-law in source 17 (NGC 4559 X7) and a single
power-law in source 18 (NGC 4559 X10), are not repeated in our results.
We examined the same \textit{XMM} observation and found multiple
background flares distributed across the whole exposure.  If we only
extract light curves at clean intervals, even including several small
flares and with a duration up to 80\% of the whole exposure, no timing
noise is present above the Poisson level.  If we use the whole exposure
including the background flares, we can reproduce the PSD reported by 
\citet{cro04}.  We conclude that the features in the power spectra
reported by \citet{cro04} are caused by background flares.

\textbf{No. 7 (NGC 3031)} We calculated the PSD from a 40.6~ks good
time interval without timing gaps or background flares for both the
MOS2 and the pn CCDs in the first half of the observation.  The second
half of the observation contains strong background flares and was
discarded.  The PSD is calculated in the energy range 2--10 keV with a
4096-point FFT.  This energy band was selected because the timing noise
is strongest in this band.  For sources 22 and 25, we also present the
PSD for the energy which maximizes the timing noise.  The time
resolution of the light curve is set to the interval duration divided by
the number of FFT points (as for all these three sources).  Power at
frequencies below $\sim0.1$ mHz increases at lower frequencies with
power-law slope $\alpha = -1.8 \pm 0.3$.  The PSD below 2 keV only has
the lowest frequency point above the Poisson level.

\textbf{No. 22 (NGC 5128)} We select a 11.4~ks good time interval
without gaps or background contamination for three MOS and pn CCDs. The
PSD is calculated in the energy range 0.7--10 keV with a 1024-point
FFT.  We also examined a power spectrum for events with energies below
0.7~keV, but there is no detectable power.  The power spectrum below
0.2~mHz has a power-law slope $\alpha = -1.56 \pm 0.14$.  There is also
some power at frequencies around 2~mHz, but our sensitivity is
inadequate to make unambiguous statements about this.

\textbf{No. 25 (NGC 5408)} The observation contains strong background
flares in the second half and a 2.96 ks good time interval in the first
half, which is used for calculation of the PSD.  \citet{sor04}  report
a power spectrum for this observation with a power-law at high frequency
breaking to a flat continuum ($\alpha = 0$) below a cutoff frequency of
2.5 mHz.  Such a PSD is often found in Galactic black hole binaries and
AGNs and the break frequency scales with mass, although the break
frequency in individual sources varies by factors of at least 10.  We
re-examine the PSD in the 0.2--10 keV band with a 1024-point FFT.  At
frequencies below 0.001~Hz the power spectrum has a power-law slope
$\alpha = -0.8 \pm 0.2$.  Above this frequency, there appears to be
power distributed irregularly across a range associated with the
flaring time scale reported by \citet{sor04}.  Our power spectrum does
not suggest a break to a flat continuum with $\alpha=0$ at low
frequency.  We are able to reproduce the spectrum reported by
\citet{sor04} if we use much broader frequency bins.

\section{Discussion}\label{sec:discussion}

We have found a diversity of spectra in our sample of ULXs. 
Considering only those sources for which the statistics justify the use
of models more complex than a simple power-law, half the sources are
best fitted with a MCD plus power-law model with the power-law dominant
at high energies.  This model is commonly used in fitting the spectra
of accreting black holes.  These sources have the lowest thermal
component temperatures, 0.1--0.4~keV, and extend to the highest
luminosities, above $10^{40}$~erg/s.  Three sources have spectra which
are strongly curved at high energies.  These sources are best fitted
with a MCD plus power-law model with the MCD component dominant at high
energies and a steep power-law at low energies or a Comptonization
model.  These sources have the highest temperatures in our sample,
1.0--1.4~keV.  Three sources show evidence of diffuse hot plasma and
are best fitted with a mekal model plus a power-law.  The temperatures
of these three sources are in the range 0.6--0.8~keV and the
luminosities are the lowest in our sample, near $10^{39}$~erg/s.
Finally, there is a single source best fitted with an MCD model alone,
This source is likely a stellar-mass black hole in a high state. 
Within each class, the spectral parameters, particularly  the emission
temperature of the thermal component, appear to be rather similar. 
Between classes, the temperature of the thermal component varies
significantly. This diversity of spectral properties and the fact that
the sources appear to lie in three distinct classes, see
Fig.~\ref{fig:lt}, suggests that the ULXs are, in fact, a diverse 
population.  

The three apparently distinct classes of ULXs may merely represent
three states of ULXs and not a true diversity of physical objects.  To
examine if this is the case, we searched the literature for state
transitions of ULXs in which the temperature of the thermal component
has been measured.  We found four such cases: Holmberg II X-1
\citep{dew04}, Holmberg IX X-1 \citet{lap01,wan04}, and two ULXs in IC
342 \citep{kub01}.  For Holmberg II and Holmberg IX X-1, multiple
observations with \textit{XMM-Newton} are available.  We plot the
spectral states of these sources, connected by dashed lines, in a
bottom panel of Fig.~\ref{fig:lt}.  As can be seen from inspection of
the figure, the ULXs which undergo state transitions remain within the
same group in the luminosity vs temperature diagram.  The state
transitions do not move the  sources between the groups.  Both sources
remain in the low $kT$ throughout the state transitions.  For the state
transitions of the two ULXs in IC 342 reported by \citet{kub01}, both
sources remain in the high $kT$ group throughout the transitions. These
sources are not plotted in  Fig.~\ref{fig:lt} because we prefer to plot
only spectra from \textit{XMM-Newton} observations.

We note that \citet{lap01} reported the state transition of M81 X-9 =
Ho IX X-1 (source 8) with a high state observed by \textit{ASCA}
fitting by a single MCD model with a high temperature $kT=1.24$ keV and
a low state fitted by a power-law plus MCD model with $kT \sim 0.3 \rm
\, keV$.  If correct, this would represent a transition between the low
$kT$ and high $kT$ groups.  However, \citet{wan04} reported two
observations using \textit{XMM-Newton} of Ho IX X-1 (one is the same as
used in this paper).  The luminosity during the \textit{XMM-Newton}
observations is a factor of 2 higher than in the  \textit{ASCA}
observation and clearly represents the high state of the source. 
\citet{wan04} concluded a single MCD model, as used by \citet{lap01} to
fit the \textit{ASCA} data, is strongly ruled out by the high quality
\textit{XMM-Newton} data.  Instead, the spectra are better described by
a cool disk plus power-law model.  Their results are consistent with
our own which rule out application of a single MCD model to the high
state of Ho IX X-1.  Therefore, we conclude that Ho IX X-1 remains in
the low $kT$ in its high and low states.

Those ULXs which undergo state transitions remain in the same class
throughout the transition, suggesting that the definition of the
different classes is robust.  We note that, on average, the low
temperature ULXs are more luminous than the high temperature ones.
Therefore, if the different classes represent different states of a
single underlying population, then the lower luminosity states would be
associated with higher temperatures, which is the reverse of the 
normal situation for black hole X-ray binaries.  However, additional
\textit{XMM-Newton} observations of the sources in our sample would be
of great interest to test if these source classes truly represent
distinct classes of objects.

\subsection{MCD plus power-law sources}

Here, we consider the properties of the class of sources which are best
fitted with MCD plus power-law model with the power-law dominant at high
energies.  In this class, there is an absence of sources with hard
spectra at luminosities near $10^{39}$~erg/s.  This is not the case for
sources which lack a detectable thermal component.  When a black hole
is in its low/hard state, the disk temperature is low \citep{mcc04}. 
If these ULXs are intermediate mass black holes (IMBHs), then their
temperature in low/hard states would shift out of the detection
threshold of \textit{XMM}.  This would naturally produce an absence of
sources with hard spectra at (relatively) low luminosities.  This
scenario would imply that the sources with single power-law spectra with
hard spectra ($\Gamma \lesssim 2$) and luminosities near $10^{39}$ erg/s 
may be IMBHs in the low/hard state.

For accreting black holes in the high state, higher black hole masses
should lead to cooler disk temperatures \citep{mak00}.  The low
temperatures of some ULXs have been suggested to be evidence that they
contain IMBHs \citep{col99,kaa03,mil03}.  In Fig.~\ref{fig:lt}, the
sources which are best fitted with a MCD plus power-law model with the
power-law dominant at high energies concentrate at low temperatures, $kT
< 0.4$~keV.  \citet{mil04a} have suggested that the clustering of  ULXs
in a narrow range in temperature is evidence that those ULXs are IMBHs
with masses lying in a narrow range.  However, current X-ray instrument
lack the sensitivity to detect thermal components with temperatures
much below 0.1~keV.  Hence, the lower boundary of the range is
observational.  The thermal emission from higher mass black holes would
simply not be detected.

In stellar mass black holes, different spectral states are associated
with different levels of variability.  The total noise power integrated
in the 0.1--10~Hz band has an rms $\lesssim 0.06$ for high states and
an rms of 0.1 to 0.3 for low states \citep{mcc04}.  Because these
relatively high frequencies are not accessible for ULXs given the
sensitivity of \textit{XMM-Newton}, we use the 0.001--0.1~Hz band to
calculate the rms noise power.  The relations between the rms noise
level and the luminosity, thermal component temperature, and power-law
photon index are shown in Fig.~\ref{fig:rms}. For those three sources
with detected rms, the noise does not appear correlated with
luminosity.  Also, the noise level does not appear correlated with
photon index.  Indeed, the one source which has a high noise level, as
associated with the hard state of stellar-mass X-ray binaries, has a
relatively soft photon index.

The X-ray variability of Galactic X-ray binaries at low frequencies has
been examined by \citet{gil05}.  In the frequency region from $10^{-4}$
to 0.1 Hz, the PSD of X-ray binaries is found to follow a power-law with
an exponent $\alpha$ in the range $-$1.2 to $-$1.5.  The PSDs for source
7, 22 and 25 at low frequencies follow power-law forms with exponents
$\alpha$ of $-1.8 \pm 0.3$, $-1.56 \pm 0.14$, and $0.8 \pm 0.2$,
respectively.  These are similar to the PSDs of Galactic X-ray binaries
at low frequencies.  This may suggest that the PSDs reflect viscous
variabilities from the accretion disks of the ULXs.

\subsection{Mekal sources\label{sec:mekal}}

The sources 12, 22 and 26 are not adequately fitted with a MCD plus
power-law model.  Their spectra present a prominent bump around
0.5--1.0 keV and their spectra are adequately fitted with the sum of a
mekal thermal plasma model plus a power-law.  The presence of a mekal
component suggests that part of the X-ray emission from these sources
arises from a hot, diffuse thermal plasem.  The fractions of the total
luminosity in the mekal component are 9.3\%, 32.2\% and 31.7\% for
source 12, 22 and 26, respectively.   

We can place an upper bound on the physical size of these sources from
that fact that all three sources appear unresolved with
\textit{Chandra}: Obs ID 402 for source 12, \citet{kra01} for source
22, and \citet{swa04} for source 26.  Therefore their angular size must
be less than 0\farcs5.  With this bound on the physical size and the
mekal fit parameters temperature $kT$ and normalization $K$, we can
then derive a lower limit on the electron density $n \ge \sqrt{4 \pi
D^2 K/(10^{14}V)}$ (cm$^{-3}$), an upper limit on the total energy of
$E \le 2 \times \case{3}{2} kT n V$ (erg), and an upper limit on the
cooling time of $ t \le 1.05 \times 10^{7} \sqrt{kT}/n$ (yr), where $D$
is the distance to the host galaxy in cm, $V$ is upper limit of the
plasma volume $V \le \case{4}{3}\pi r^3$ in cm$^3$, $kT$ is the plasma
temperature in keV, and the value and units of $K$ are taken from
\textsl{XSPEC}.

\textbf{No. 12 (NGC 4395)} The source is also known as NGC 4395 X-2
\citep{lir00} located in the dwarf galaxy NGC 4395. In the \textit{XMM}
observation on 2003-Nov-30, the source is located near the gap in pn
CCDs, leading to a small difference in flux between pn and MOS CCDs.
Therefore we only take the two MOS CCDs for spectral analysis for this
source. The spectral fitting gives $kT = 0.78 \pm 0.04$~keV and $K =
4.5 \pm 0.5 \times 10^{-5}$, from which we derive $n \geq
9.3$ cm$^{-3}$, $E \leq 2.8 \times 10^{51}$~erg and $ t \leq 1.0 \times
10^6$~yr.

\textbf{No. 22 (NGC 5128)} The source is located near the edge of
southwest radio lobe in NGC 5128 (Centaurus A).  Our result shows the
source has a strong mekal component.  The derived
plasma properties from the temperature $kT = 0.68 \pm 0.03$~keV and
normalization $K = 4.1 \pm 0.04 \times 10^{-5}$ are $n \geq
7.6$~cm$^{-3}$, $E \leq 5.1 \times 10^{51}$~erg and $t \leq 1.1 \times
10^6$~yr. The X-ray source, according to the coordinates RA=13h25m07.45s, 
Dec=-43d04m09.53s (J2000) from Chandra \citep{kra01}, corresponds to an 
optical counterpart that is included in the USNO-B1.0 catalog \citep{mon03}
with an apparent magnitude $m_{B2}=14.23$ mag and $m_{R2}=14.41$ mag, 
corresponding to an absolute magnitude $M_{B2}=-14.2$ mag and $M_{R2}=-14.0$.
The B$-$R color of the source is inconsistent with a foreground M dwarf 
as suggested by \citep{kra01}.

\textbf{No. 26 (NGC 5457)} The source is located $6\arcmin$ west of the
nucleus of NGC 5457 (M101), with a relatively low luminosity
($L<10^{39}$~erg/s for the best-fitting model) in this \textit{XMM}
observation.  The mekal component is prominent in this observation with
spectral parameters $kT = 0.67 \pm 0.04$~keV and $K=1.7\pm0.2 \times
10^{-5}$.  We derive the plasma parameters $n \geq 4.0$~cm$^{-3}$, $E
\leq 7.8 \times 10^{51}$~erg and $t \leq 2.1 \times 10^6$~yr.  This
source also has an optical counterpart in the DSS image.  The optical
counterpart is in the GSC catalog \citep{mor01} and has an  apparent
magnitude $m_\mathrm{V} = 15.5 \pm 0.3$~mag, corresponding to an
absolute magnitude $M_\mathrm{V}=-13.7$~mag.  This source is also
observed by the Optical Monitor on-aboard \textit{XMM} with an absolute
magnitude $M_\mathrm{UVW1}=-10.8$~mag.

One plausible interpretation for the hot diffuse plasma components is
that they are young supernova remnants.  The derived density, total
energy, and cooling time are consistent with this interpretation.  The
X-ray luminosities and temperatures of these three sources are similar
to recent supernovae \citep[e.g.\ SN 1979C,][]{imm03,imm05}.  The high
absolute magnitudes of the optical counterparts indicates they are not
normal stars.  Comparison with the late-time absolute magnitude of SN
1979C ($M_\mathrm{B}=-12.1$~mag, $M_\mathrm{UVW1}=-13.6$~mag), suggests
that they may be supernova remnants which are several tens years old.
We do not find supernovae counterparts on the IAU supernova
list\footnote{\url{http://cfa-www.harvard.edu/iau/lists/Supernovae.html}},
but the supernova may easily have been missed.  Source 22 is found to
have noise power above the Poisson level, which could not be produced
by a spatially extended source.  However, more than 2/3 of the
luminosity arises from the power-law emission.  The low frequency noise
for source 22 (the first two points in Fig.~\ref{fig:psd}) is strongest
in the 1--10 keV band, where the energy spectrum is dominated by the
power-law component also.  Therefore, the variability appears related
to the power-law rather than the mekal component and may arise from a 
central compact object.

\subsection{High temperature sources\label{sec:comp}}

Three sources (3, 6, and 7) have spectra which are strongly curved at
high energies in strong contrast to most of the other sources.  From
the MCD fits, we found that the sources have high temperatures and are
either super-Eddington for a stationary black hole or near Eddington
for a rapidly rotating black hole.  These unusual aspects establish
these sources as a distinct class. All three sources have low timing
noise with an rms in the  0.001--0.1~Hz band below 0.02.

\citet{fos04} suggested that source 3 (M33 X-8) is a stellar-mass black
hole in the high state because its behavior is similar to that of
Galactic black hole X-ray binaries.  Radio emission with a spectral
index similar to microquasars has been found in VLA observations
\citep{gor99}, possibly linking this source with microquasars. 
\citet{fos04} also suggest that the emission may be beamed.  Source 7
(in M81, F88 X-6) has also been identified with a weak radio
counterpart \citep{fab88,swa03}.

Other interpretations for the high temperatures in some ULXs include
intermediate mass black holes in the Kerr metric \citep{mak00} or 
black holes of several tens of solar masses with slim accretion disks
(advection dominated optically thick disks) emitting at super-Eddington
luminosities \citep{ebi03}.  These interpretations are based on results
showing that single multicolor disk blackbody spectrum provide the best
fits for high-temperature ULXs \citep{mak00}.  Our results based on the
high quality spectra available from \textit{XMM-Newton} show that a
single MCD provides a significantly worse fit than a Comptonization
model.  A MCD model with an additional power-law component can provide
an equally good fit, but the power-law is dominant at low energies,
which is different from the behavior usually seen from Galactic black
hole binaries where the power-law component is usually weak when then
source is in the thermal dominant state which shows prominent MCD
emission near 1~keV \citep{mcc04}.  Interpretation of the low-energy 
part of power-law as due to Comptonization would be problematic because 
the seed photons upscattered by Comptonization are usually assumed to 
arise from the disk emission \citep{rob05}.  Power-law components extending 
below the disk emission have been found in the spectra of the black 
hole candidates LMC X-1 and LMC X-3 \citep{haa01}, but are likely better interpreted
in terms of physically motivated Comptonized MCD models \citep{yao05}.
In the Comptonization model, a high
electron column density $\sim 3 \times 10^{25} \rm \, cm^{-2}$, is
required to produce the observed optical depths \citep{mak00}.  Since
heavy ions would not be ionized at the inferred temperatures, strong
Fe K absorption should be observed unless the material has very low
metallicity.  No Fe K absorption is apparent in the data, but the
constraints are weak due to the limited statistics at high energies. 
Adding a smeared edge model to the Comptonization spectra with a
central energy of 7~keV, optical depths up to $\tau \sim 6$ are
allowed, depending on the edge width.  Lack of Fe K absorption would
argue against accepting the Comptonization model.

\acknowledgments  

We thank the referee and R. Soria for some useful suggestions which improved
our paper. We acknowledge partial support from Chandra grants GO4-5086 and
GO4-5035.  PK thanks the Aspen Center for Physics for its hospitality
during the workshop, ``Compact Objects in External Galaxies'', and
acknowledges partial support from a University of Iowa Faculty Scholar
Award.

\begin{deluxetable*}{rlllcrcc}[t]
\tabletypesize{\footnotesize}
\tablewidth{0pc}
\tablecaption{XMM Observations of ULXs in nearby galaxies\label{tab:data}}
\tablehead{
  \colhead{No.} & \colhead{Host Galaxy} & \colhead{RA} & \colhead{Dec} &
  \colhead{Dist.} & \colhead{Obs Date} & \colhead{Exposure\tablenotemark{a}} & \colhead{Ref.\tablenotemark{b}}\\
  \colhead{} & \colhead{} & \colhead{(J2000)} & \colhead{(J2000)} &
  \colhead{(Mpc)} & \colhead{} & \colhead{(ks)} & \colhead{}
}
\startdata
1 & Circinus & 14 12 53.52 & $-$65 22 54.7 & 4.0 & 08/06/01 & 61.1/91.1 & 1\\
2 & IC 342 & 03 45 55.7 & $+$68 04 58 & 3.9 & 02/11/01 & 4.9/5.1 & 2 \\
3 & NGC 598 & 01 33 50.89 & $+$30 39 37.2 & 0.9 & 08/04/00 & 6.5/11.9 & 1 \\
4 & NGC 1313 & 03 18 19.9 & $-$66 29 07 & 3.7 & 10/17/00 & 17.4/28.0 & 2\\
5 & NGC 1313 & 03 18 22.2 & $-$66 36 01 & 3.7 & 10/17/00 & 17.4/28.0 & 2\\
6 & NGC 2403 & 07 36 25.55 & $+$65 35 40.0 & 3.1 & 09/12/04 & 50.6/69.1 & 1\\
7 & NGC 3031 & 09 55 32.98 & $+$69 00 33.4 & 3.6 & 04/22/01 & 87.8/125.4 & 1\\
8 & NGC 3031 & 09 57 54.4 & $+$69 03 43 & 3.6 & 04/16/02 & 7.7/7.7 & 2\\
9 & NGC 3628 & 11 20 15.76 & $+$13 35 13.7 & 10.0 & 11/27/00 & 37.7/41.2 & 1\\
10 & NGC 4258 & 12 18 43.86 & $+$47 17 31.5 & 7.7 & 12/08/00 & 12.2/14.4 & 1\\
11 & NGC 4258 & 12 18 57.84 & $+$47 16 07.1 & 7.7 & 12/08/00 & 12.2/14.4 & 1\\
12 & NGC 4395 & 12 26 01.8 & $+$33 31 34 & 3.6 & 11/30/03 & 86.9/96.1 & 2\\
13 & NGC 4449 & 12 28 17.83 & $+$44 06 33.9 & 3.7 & 06/02/02 & 10.6/11.8 & 1\\
14 & NGC 4485 & 12 30 30.56 & $+$41 41 42.3 & 7.8 & 05/27/02 & 11.5/12.0 & 1\\
15 & NGC 4490 & 12 30 36.32 & $+$41 38 37.8 & 7.8 & 05/27/02 & 11.5/12.0 & 1\\
16 & NGC 4490 & 12 30 43.25 & $+$41 38 18.4 & 7.8 & 05/27/02 & 11.5/12.0 & 1\\
17 & NGC 4559 & 12 35 51.71 & $+$27 56 04.1 & 10.3 & 05/27/03 & 33.7/36.4 & 1\\
18 & NGC 4559 & 12 35 58.57 & $+$27 57 41.8 & 10.3 & 05/27/03 & 33.7/36.4 & 1\\
19 & NGC 4631 & 12 41 55.56 & $+$32 32 16.9 & 7.6 & 06/28/02 & 44.9/53.2 & 1\\
20 & NGC 4945 & 13 05 21.94 & $-$49 28 26.6 & 3.7 & 01/21/01 & 17.9/17.9 & 1\\
21 & NGC 4945 & 13 05 32.89 & $-$49 27 34.1 & 3.7 & 01/21/01 & 17.9/17.9 & 2\\
22 & NGC 5128 & 13 25 07.4 & $-$43 04 06 & 4.9 & 02/02/01 & 16.8/16.8 & 2\\
23 & NGC 5204 & 13 29 38.61 & $+$58 25 05.6 & 4.5 & 01/06/03 & 15.4/15.4 & 1\\
24 & NGC 5236 & 13 37 20.1 & $-$29 53 46 & 4.7 & 01/27/03 & 18.9/24.1 & 2\\
25 & NGC 5408 & 14 03 19.61 & $-$41 22 59.6 & 4.8 & 01/28/03 & 2.7/4.0 & 3\\
26 & NGC 5457 & 14 02 29.89 & $+$54 21 19.0 & 7.0 & 06/04/02 & 21.6/36.5 & 1\\
27 & NGC 5457 & 14 04 14.28 & $+$54 26 03.6 & 7.0 & 06/04/02 & 21.6/36.5 & 1\\
28 & UGC 4305 & 08 19 28.99 & $+$70 42 19.4 & 3.4 & 04/10/02 & 4.7/4.7 & 1\\
\enddata

\tablenotetext{a}{The clean/total exposure of pn CCDs, or mos CCDs in
case pn data are not used.}
\tablenotetext{b}{The catalog from which the coordinates of the ULXs
and the distance of the host galaxy are taken; (1)--\citet{swa04},
(2)--\citet{col02}, (3)--\citet{liu05}.}
\end{deluxetable*}

\clearpage
\begin{deluxetable}{rllllllcl}
\tabletypesize{\footnotesize}
\tablewidth{0pc}
\tablecaption{Spectral parameters and variabilities of selected ULXs.\label{tab:spec}}
\tablehead{
  \colhead{No.} & \colhead{Model} & \colhead{$n'_\mathrm{H}$\tablenotemark{a}} & \colhead{$n_\mathrm{H}$} & \colhead{$\Gamma$/$\tau$\tablenotemark{b}} & \colhead{$kT$\tablenotemark{c}} &\colhead{$L$\tablenotemark{d}} & \colhead{$\chi^2/$dof} & \colhead{rms\tablenotemark{e}}\\
  \colhead{} & \colhead{} & \colhead{($10^{21}$ cm$^{-2}$)} & \colhead{($10^{21}$ cm$^{-2}$)} &\colhead{} & \colhead{(keV)} & \colhead{($10^{40}$ erg/s)} 
  & \colhead{} & \colhead{(0.001--0.1 Hz)}
}
\startdata
1 & powerlaw & 5.33 & $10.9_{-0.7}^{+0.7}$ & $3.30_{-0.12}^{+0.12}$ & \nodata &
$0.42_{-0.14}^{+0.01}$ & 340.4/265 & $<0.131$\\
& diskbb &  & $5.6_{-0.4}^{+0.4}$ & \nodata & $0.74_{-0.03}^{+0.03}$
 & $0.0764_{-0.0067}^{+0.0002}$ & 267.4/265 & \\
2 & powerlaw & 3.03 & $5.8_{-0.6}^{+0.6}$ & $1.67_{-0.05}^{+0.09}$ & \nodata &
$0.44_{-0.05}^{+0.04}$ & 142.9/153 & $<0.069$\\
& diskbb & & 
$3.1_{-0.1}^{+0.4}$ & \nodata & $2.01_{-0.15}^{+0.17}$ & $0.48_{-0.04}^{+0.01}$
 & 169.2/153 & \\
3 & diskbb & 0.56 & 
$0.56_{+0.00}^{+0.02}$ & \nodata & $1.12_{-0.02}^{+0.01}$ & 
$0.156_{-0.003}^{+0.001}$ & 909.0/692 & $<0.003$\\
& powerlaw+diskbb &  & $1.7_{-0.2}^{+0.2}$ & $2.34_{-0.13}^{+0.16}$ & 
$1.14_{-0.06}^{+0.06}$ & $0.226_{-0.003}^{+0.006}$ & 694.1/690 & \\
& compst & & $1.6_{-0.1}^{+0.1}$ & $20.5_{-0.8}^{+0.9}$ & $1.18_{-0.05}^{+0.05}$
 & $0.220_{-0.003}^{+0.003}$ & 700.1/691&\\
4 & powerlaw & 0.39 & $1.9_{-0.1}^{+0.1}$ & $1.95_{-0.04}^{+0.04}$ & \nodata &
$0.59_{-0.01}^{+0.02}$ & 766.2/627 & $0.028 \pm 0.013$\\
& powerlaw+diskbb &  & $3.0_{-0.3}^{+0.4}$ & $1.74_{-0.06}^{+0.06}$ & 
$0.20_{-0.02}^{+0.02}$ & $0.81_{-0.06}^{+0.03}$ & 598.9/625 & \\
5 & powerlaw & 0.39 & $2.7_{-0.2}^{+0.2}$ & $2.45_{-0.08}^{+0.08}$ & \nodata &
$0.29_{-0.01}^{+0.01}$ & 266.9/252 & $<0.077$\\
& powerlaw+diskbb &  & $2.9_{-0.3}^{+0.7}$ & $2.19_{-0.15}^{+0.18}$ & 
$0.23_{-0.06}^{+0.05}$ & $0.30_{-0.05}^{+0.02}$ & 244.8/250 & \\
6 & diskbb & 0.41 & 
$1.6_{-0.1}^{+0.1}$ & \nodata & $1.04_{-0.02}^{+0.02}$ & 
$0.170_{-0.003}^{+0.001}$ & 1019.2/869 & $<0.012$\\
& powerlaw+diskbb &  & $4.4_{-0.7}^{+1.0}$ & $3.5_{-0.5}^{+0.7}$ & 
$1.16_{-0.04}^{+0.04}$ & $0.42_{-0.04}^{+0.01}$ & 952.3/867 & \\
& compst & & $3.0_{-0.2}^{+0.1}$ & $21.5_{-1.1}^{+1.3}$ & $1.04_{-0.05}^{+0.05}$
 & $0.232_{-0.005}^{+0.007}$ & 942.6/868&\\
7 & diskbb & 0.42 & 
$1.3_{-0.1}^{+0.1}$ & \nodata & $1.29_{-0.03}^{+0.03}$ & $0.43_{-0.01}^{+0.01}$
 & 399.9/310 & $<0.008$\\
& powerlaw+diskbb &  & $4.0_{-0.7}^{+1.4}$ & $3.2_{-0.5}^{+1.0}$ & 
$1.51_{-0.08}^{+0.06}$ & $0.87_{-0.07}^{+0.02}$ & 317.7/308 & \\
& compst & & $2.7_{-0.2}^{+0.1}$ & $20.3_{-1.3}^{+1.2}$ & $1.31_{-0.07}^{+0.08}$
 & $0.56_{-0.02}^{+0.02}$ & 315.8/309&\\
8 & powerlaw & 0.42 & $2.0_{-0.1}^{+0.1}$ & $1.88_{-0.03}^{+0.03}$ & \nodata &
$1.69_{-0.03}^{+0.03}$ & 979.3/944 & $<0.010$\\
& powerlaw+diskbb &  & $2.4_{-0.2}^{+0.2}$ & $1.81_{-0.05}^{+0.04}$ & 
$0.22_{-0.03}^{+0.05}$ & $1.85_{-0.11}^{+0.05}$ & 942.0/942 & \\
9 & powerlaw & 0.22 & $6.6_{-0.4}^{+0.5}$ & $1.65_{-0.06}^{+0.07}$ & \nodata &
$1.06_{-0.05}^{+0.05}$ & 296.5/280 & $<0.061$\\
10 & powerlaw & 0.12 & $1.3_{-0.4}^{+0.5}$ & $1.8_{-0.2}^{+0.2}$ & \nodata &
$0.13_{-0.01}^{+0.02}$ & 43.8/40 & $<0.498$\\
11 & powerlaw & 0.12 & $2.6_{-0.5}^{+0.7}$ & $2.00_{-0.17}^{+0.10}$ & \nodata &
$0.16_{-0.02}^{+0.02}$ & 48.1/46 & $<0.253$\\
& diskbb & & 
$0.7_{-0.3}^{+0.4}$ & \nodata & $1.20_{-0.16}^{+0.19}$ & $0.10_{-0.02}^{+0.01}$
 & 55.8/46 & \\
12
& powerlaw+mekal & 0.14 & $1.6_{-0.1}^{+0.1}$ & $3.73_{-0.10}^{+0.11}$ & 
$0.78_{-0.04}^{+0.04}$ & $0.20_{-0.01}^{+0.01}$ & 
292.6/217 & $<0.045$\\
& powerlaw+diskbb & & $1.1_{-0.2}^{+0.3}$ & $3.1_{-0.2}^{+0.3}$ & 
$0.25_{-0.02}^{+0.01}$ & $0.120_{-0.002}^{+0.022}$ & 375.7/217&\\
13 & powerlaw & 0.14 & $6.5_{-0.9}^{+1.0}$ & $2.25_{-0.08}^{+0.16}$ & \nodata &
$0.19_{-0.03}^{+0.01}$ & 80.9/90 & $<0.073$\\
& diskbb & & 
$2.7_{-0.6}^{+0.6}$ & \nodata & $1.26_{-0.11}^{+0.12}$ & 
$0.096_{-0.014}^{+0.001}$ & 112.4/90 & \\
14 & powerlaw & 0.18 & $2.6_{-0.5}^{+0.5}$ & $2.06_{-0.15}^{+0.17}$ & \nodata &
$0.50_{-0.04}^{+0.05}$ & 64.3/57 & $<0.092$\\
& diskbb & & 
$0.8_{-0.3}^{+0.3}$ & \nodata & $1.12_{-0.11}^{+0.12}$ & $0.29_{-0.04}^{+0.01}$
 & 74.5/57 & \\
15 & powerlaw & 0.18 & $3.2_{-0.3}^{+0.3}$ & $2.08_{-0.09}^{+0.09}$ & \nodata &
$0.63_{-0.04}^{+0.04}$ & 145.4/142 & $<0.077$\\
& diskbb & & 
$1.1_{-0.2}^{+0.2}$ & \nodata & $1.25_{-0.07}^{+0.08}$ & $0.38_{-0.03}^{+0.01}$
 & 159.6/142 & \\
16 & powerlaw & 0.18 & $6.8_{-0.8}^{+0.9}$ & $2.59_{-0.15}^{+0.16}$ & \nodata &
$0.79_{-0.17}^{+0.04}$ & 97.2/89 & $<0.132$\\
& diskbb & & 
$3.3_{-0.5}^{+0.5}$ & \nodata & $0.98_{-0.07}^{+0.07}$ & $0.31_{-0.04}^{+0.01}$
 & 94.8/89 & \\
17 & powerlaw & 0.15 & $1.3_{-0.1}^{+0.1}$ & $2.39_{-0.05}^{+0.06}$ & \nodata &
$1.55_{-0.04}^{+0.06}$ & 442.2/360 & $<0.038$\\
& powerlaw+diskbb &  & $1.8_{-0.2}^{+0.4}$ & $2.13_{-0.07}^{+0.09}$ & 
$0.17_{-0.03}^{+0.01}$ & $1.94_{-0.13}^{+0.11}$ & 353.0/358 & \\
18 & powerlaw & 0.15 & $1.2_{-0.1}^{+0.1}$ & $1.97_{-0.05}^{+0.06}$ & \nodata &
$1.30_{-0.05}^{+0.06}$ & 238.6/256 & $<0.061$\\
19 & powerlaw & 0.13 & $2.7_{-0.3}^{+0.3}$ & $2.07_{-0.10}^{+0.10}$ & \nodata &
$0.51_{-0.03}^{+0.03}$ & 127.1/128 & $<0.049$\\
& diskbb & & 
$0.7_{-0.2}^{+0.2}$ & \nodata & $1.20_{-0.08}^{+0.08}$ & $0.30_{-0.02}^{+0.01}$
 & 159.3/128 & \\
20 & powerlaw & 1.55 & $5.3_{-0.8}^{+0.9}$ & $2.5_{-0.2}^{+0.2}$ & \nodata &
$0.08_{-0.02}^{+0.01}$ & 73.8/59 & $<0.160$\\
& diskbb & & 
$2.7_{-0.5}^{+0.6}$ & \nodata & $0.85_{-0.08}^{+0.09}$ & 
$0.033_{-0.006}^{+0.001}$ & 67.1/59 & \\
21 & powerlaw & 1.54 & $5.1_{-0.6}^{+0.7}$ & $1.76_{-0.09}^{+0.10}$ & \nodata &
$0.13_{-0.01}^{+0.01}$ & 167.0/155 & $<0.073$\\
& diskbb & & 
$2.7_{-0.4}^{+0.4}$ & \nodata & $1.69_{-0.12}^{+0.13}$ & $0.09_{-0.01}^{+0.01}$
 & 151.3/155 & \\
22
& powerlaw+mekal & 0.87 & $0.87_{+0.00}^{+0.19}$ & $3.01_{-0.11}^{+0.19}$ & 
$0.68_{-0.03}^{+0.03}$ & $0.066_{-0.008}^{+0.004}$ & 179.0/147 & $<0.069$\\
& powerlaw+diskbb & & $1.1_{-0.2}^{+1.0}$ & $2.1_{-1.3}^{+1.6}$ & 
$0.26_{-0.06}^{+0.01}$ & $0.11_{-0.01}^{+0.54}$ & 266.1/147&\\
23 & powerlaw & 0.14 & $0.6_{-0.1}^{+0.1}$ & $2.05_{-0.04}^{+0.04}$ & \nodata &
$0.46_{-0.01}^{+0.01}$ & 606.2/574 & $0.011 \pm 0.005$\\
& powerlaw+diskbb &  & $0.4_{-0.1}^{+0.1}$ & $1.77_{-0.10}^{+0.09}$ & 
$0.31_{-0.04}^{+0.04}$ & $0.45_{-0.02}^{+0.05}$ & 571.1/572 & \\
24 & powerlaw & 0.37 & $1.5_{-0.2}^{+0.2}$ & $2.45_{-0.10}^{+0.10}$ & \nodata &
$0.17_{-0.01}^{+0.01}$ & 267.5/236 & $<0.045$\\
25 & powerlaw & 0.56 & $1.4_{-0.2}^{+0.2}$ & $3.06_{-0.12}^{+0.14}$ & \nodata &
$0.90_{-0.04}^{+0.04}$ & 218.4/171 & $0.100 \pm 0.010$\\
& powerlaw+diskbb &  & $1.1_{-0.2}^{+0.2}$ & $2.3_{-0.2}^{+0.3}$ & 
$0.21_{-0.03}^{+0.03}$ & $0.73_{-0.06}^{+0.10}$ & 177.9/169 & \\
26
& powerlaw+mekal & 0.11 & $0.4_{-0.2}^{+0.2}$ & $2.6_{-0.1}^{+0.3}$ & 
$0.67_{-0.04}^{+0.04}$ & $0.08_{-0.01}^{+0.01}$ & 115.5/86 & $<0.465$\\
& powerlaw+diskbb & & $1.4_{-0.9}^{+0.9}$ & $1.3_{-1.4}^{+0.8}$ & 
$0.23_{-0.04}^{+0.03}$ & $0.181_{-0.077}^{+0.005}$ & 160.7/86&\\
27 & powerlaw & 0.12 & $1.6_{-0.3}^{+0.3}$ & $2.8_{-0.2}^{+0.2}$ & \nodata &
$0.28_{-0.02}^{+0.02}$ & 101.0/77 & $<0.336$\\
28 & powerlaw & 0.34 & $1.8_{-0.1}^{+0.1}$ & $2.73_{-0.03}^{+0.04}$ & \nodata &
$1.85_{-0.03}^{+0.03}$ & 892.8/731 & $<0.004$\\
& powerlaw+diskbb &  & $1.5_{-0.2}^{+0.2}$ & $2.45_{-0.11}^{+0.11}$ & 
$0.31_{-0.04}^{+0.04}$ & $1.52_{-0.03}^{+0.11}$ & 864.6/729 & \\

\enddata
\tablenotetext{a}{Absorption column density of the Milky Way \citep{dic90}}
\tablenotetext{b}{$\Gamma$ means the photon index for models including
a power-law component; $\tau$ means the optical depth for the compst
model.}
\tablenotetext{c}{$kT$ stands for the inner disk temperature for the
diskbb model, the plasma temperature for the mekal model, and the
electron temperature for the compst model.}
\tablenotetext{d}{Unabsorbed X-ray luminosity in the energy range
0.3--10 keV.}
\tablenotetext{e}{rms variabilities in the frequency range 0.001--0.1
Hz. The rms values are not associated with the
spectral models, while in Fig. \ref{fig:rms} the luminosity is selected
from the best-fitted model.}
\end{deluxetable}


\begin{thebibliography}{}

\bibitem[Colbert \& Mushotzky(1999)]{col99} Colbert, E.J.M., \&
Mushotzky, R.F. 1999, \apj, 519, 89

\bibitem[Colbert \& Ptak(2002)]{col02} Colbert, E.J.M., \& Ptak, A.F.
2002, \apjs, 143, 25

\bibitem[Cropper et al.(2004)]{cro04} Cropper, M.\ et al.\ 2004, MNRAS,
349, 39

\bibitem[Dewangan et al.(2004)]{dew04} Dewangan, G.C. et al. 2004, \apj,
608, L57

\bibitem[Dickey \& Lockman(1990)]{dic90}
Dickey, J.M., \& Lockman, F.J. 1990, \araa, 28, 215

\bibitem[Ebisawa et al.(2003)]{ebi03}
Ebisawa, K. et al. 2003, \apj, 597, 780

\bibitem[Fabbiano(1988)]{fab88} Fabbiano, G. 1988, \apj, 325, 544

\bibitem[Fabbiano(1989)]{fab89} Fabbiano, G. 1989, \araa, 27, 87

\bibitem[Fabbiano \& White(2003)]{fab03} Fabbiano, G., \& White, N.E.
2003, Compact Stellar X-ray Sources in Normal Galaxies, eds. W.H.G.
Lewin and M. van der Klis (Cambridge: Cambridge Univ. Press), in press,
astro-ph/0307077

\bibitem[Foschini et al.(2004)]{fos04} Foschini, L. et al. 2004, \aap,
416, 529

\bibitem[Gilfanov \& Arefiev(2005)]{gil05} Gilfanov, M. \& Arefiev, V.
2005, submitted to \mnras, astro-ph/0501215

\bibitem[Gordon et al.(1999)]{gor99} Gordon, S.M. et al. 1999, \apjs,
120, 247

\bibitem[Haardt et al.(2001)]{haa01}
Haardt, F. et al. 2001, \apjs, 133, 187 

\bibitem[Immler et al.(2005)]{imm05} Immler, R. et al. 2005, submitted
to \apj, astro-ph/0503678

\bibitem[Immler \& Lewin(2003)]{imm03} Immler, R., \& Lewin, W.H.G.
2003, in Supernovae and Gamma-Ray Bursts, ed. K.W. Weiler (Springer-Verlag), 91,
astro-ph/0202231

\bibitem[Kaaret et al.(2001)]{kaa01} Kaaret, P. et al. 2001, \mnras,
321, L29

\bibitem[Kaaret et al.(2003)]{kaa03} Kaaret, P. et al. 2003, Science,
299, 365

\bibitem[Kaaret et al.(2004)]{kaa04} Kaaret, P. et al. 2004, \mnras,
351, L83

\bibitem[King et al.(2001)]{kin01} King, A.R. et al. 2001, ApJ, 552,
L109

\bibitem[K\"ording et al.(2002)]{kor02}  K\"ording, E. et al. 2002,
\aap, 382, L13

\bibitem[Kraft et al.(2001)]{kra01} Kraft, R.P. et al. 2001, \apj, 560,
675

\bibitem[Kubota et al.(2001)]{kub01} Kubota, A. et al. 2001, \apj, 547, L119


\bibitem[La Parola et al.(2001)]{lap01} La Parola, V. et al. 2001, \apj,
556, 47

\bibitem[Lira et al.(2000)]{lir00} Lira, P. et al. 2000, \mnras, 319,
17

\bibitem[Liu \& Mirabel(2005)]{liu05} Liu, Q.Z., \& Mirabel, I.F. 2005,
\aap, 429, 1125

\bibitem[Makishima et al.(1986)]{mak86} Makishima, K. et al. 1986,
\apj, 308, 635

\bibitem[Makishima et al.(2000)]{mak00} Makishima, K. et al. 2000,
\apj, 535, 632

\bibitem[McClintock \& Remillard(2004)]{mcc04} McClintock, J.E., \&
Remillard, R.A. 2004, Black Hole Binaries, eds. W.H.G. Lewin and M. van
der Klis (Cambridge: Cambridge Univ. Press), in press, astro-ph/0306213

\bibitem[Mewe et al.(1995)]{mew95} Mewe, R. et al. 1995, Legacy 6, 16

\bibitem[Miller et al.(2003)]{mil03} Miller, J.M. et al. 2003, \apj,
585, L37

\bibitem[Miller et al.(2004)]{mil04a} Miller, J.M. et al. 2004, \apj,
614, L117

\bibitem[Miller \& Colbert(2004)]{mil04b} Miller, M.C. \& Colbert,
E.J.M. 2004, Int. J. Mod. Phys., 13, 1, astro-ph/0308402

\bibitem[Mitsuda et al.(1984)]{mit84} Mitsuda, K. et al. 1984, \pasj,
36, 741

\bibitem[Monet et al.(2003)]{mon03} Monet, D.G. et al. 2003, \aj,
125, 984

\bibitem[Morrison et al.(2001)]{mor01} Morrison J.E. et al. 2001, \aj,
121, 1752

\bibitem[Pakull \& Mirioni(2003)]{pak03} Pakull, M.W., \& Mirioni, L.
2003, in ESA SP--488, New Visions of the X-ray Universe in the
XMM-Newton and Chandra Era, ed. F. Jansen (Noordwijk: ES- TEC/ESA), in
press (astro-ph/0202488)

\bibitem[Ptak \& Griffiths(1999)]{pta99} Ptak, A., \& Griffiths R.
1999, \apj, 517, L85

\bibitem[Roberts et al.(2005)]{rob05}
Roberts, T. P., Warwick, R. S., Ward, M. J., Goad, M. R., Jenkins, L.
P. 2005, \mnras, 357, 1363 

\bibitem[Soria et al.(2004)]{sor04} Soria, R. et al. 2004, \aap, 423,
955

\bibitem[Swartz et al.(2003)]{swa03} Swartz, D.A. et al. 2003, \apjs,
144, 213

\bibitem[Swartz et al.(2004)]{swa04} Swartz, D.A. et al. 2004, \apjs,
154, 519

\bibitem[Sunyaev \& Titarchuk(1980)]{sun80} Sunyaev, R.A., \&
Titarchuk, L.G. 1980, \aap, 86, 121

\bibitem[van de Klis(1988)]{van88} van der Klis, M. 1988, in Timing
Neutron Stars,  ed. H. Ogelmen \& E. P. J. van den Heuvel (Dordrecht:
Kluwer), 27

\bibitem[Wang et al.(2004)]{wan04} Wang, Q.D. et al. 2004, \apj, 609, 113

\bibitem[Yao, Wang \& Zhang(2005)]{yao05}
Yao, Y., Wang, D. Q., Zhang, S. N. 2005, \mnras, 362, 229 

\end{thebibliography}
\end{document}